\documentclass[twocolumn,twocolappendix]{aastex631}

\usepackage{appendix}
\usepackage[utf8]{inputenc}
\usepackage{graphicx}
\usepackage{float}
\usepackage{enumerate}
\usepackage{comment}
\usepackage{enumitem}
\usepackage{gensymb}

\begin{document}

\title{Measuring Dust Masses of Protoplanetary Disks in Serpens \& L1641/L1647 with ALMA}

\author[0000-0001-9219-7696]{Luisa F. Zamudio-Ruvalcaba}
\affiliation{Department of Astronomy, Boston University, 725 Commonwealth Avenue, Boston, MA 02215,
USA}
\affiliation{Institute for Astrophysical Research, Boston University, 725 Commonwealth Avenue, Boston, MA 02215,
USA}

\author[0000-0001-9227-5949]{Catherine C. Espaillat}
\affiliation{Department of Astronomy, Boston University, 725 Commonwealth Avenue, Boston, MA 02215,
USA}
\affiliation{Institute for Astrophysical Research, Boston University, 725 Commonwealth Avenue, Boston, MA 02215,
USA}

\author[0000-0003-3133-3580]{Álvaro Ribas}
\affiliation{Institute of Astronomy, University of Cambridge, Madingley Road, Cambridge, CB3 0HA, UK}

\author[0000-0003-1283-6262]{Enrique Macías}
\affiliation{European Southern Observatory, Karl-Schwarzschild-Str. 2, 85748 Garching bei München, Germany}

\begin{abstract}

Protoplanetary disks are an essential component of the planet-formation process. The amount of dust and gas in the disk constrains the number and size of planets that can form in a system. We analyze 178 T-Tauri stars, 18 in Serpens and 160 in L1641/L1647, and measure their disk dust masses using spectral energy distribution (SED) modeling and multiwavelength data, including 1.3 mm (ALMA band 6) fluxes from the literature. The disk masses calculated in this work are up to $\sim$2 times higher than those previously reported. We conclude that this is because disks may be partially optically thick at millimeter wavelengths while most calculations of the disk mass assume that the disk is optically thin at 1.3~mm. We calculate optical depths at 1.3 and 7 mm for a subset of the Serpens and L1641/L1647 disk sample and show that the vast majority of disks become optically thin at longer millimeter wavelengths; thus, observations at 7~mm (i.e., ALMA band 1) are vital to better characterize disk dust masses.

\end{abstract}

\keywords{accretion, accretion disks ---  planets and satellites: formation --- protoplanetary disks --- T Tauri stars}

\section{Introduction} \label{sec:intro}

 Stars form from collapsing molecular cloud cores \citep{1993Goodman} and, due to angular momentum conservation, develop a disk of gas and dust (i.e., a protoplanetary disk).  The total mass of the disk dictates the number of resulting planets and the timescale for the planet formation process \citep{2020Tychoniec}. A common assumption is that protoplanetary disks have a gas-to-dust ratio of 100 \citep[][]{1987Shu, 2010Birnstiel}.
 Although dust makes up only a small fraction of the disk’s mass, it dominates the disk’s opacity and tends to be more easily observed than gas. By studying the thermal emission from dust grains, we can infer the dust mass and gain insight into the conditions under which planets form.

Disk dust masses can be calculated with different methods \citep[see][for a review]{2018BerginWilliams}, but a common method is using millimeter-wavelength flux measurements. Some of the typical values measured for disk dust masses are $\sim0.2 - 0.4\ M_{\oplus}$, which are insufficient to support ongoing planet formation \citep[see][]{2016Ansdell, 2016Pascucci, 2018Manara} since it implies a planet formation efficiency rate close to 100$\%$ \citep{2021Mulders}. This is called the ``missing mass''  problem (also known as the ``mass budget'' problem), and different solutions have been proposed. Planet formation may start much earlier than previously assumed,
even during the first Myr of the disk lifetime \citep[e.g.][]{2014NajitaKenyon,2018Manara,2021Mulders}, which is supported by the detection of disk substructures \citep[e.g.][]{2015ALMApartnership,2018Andrews,2020SeguraCox} and the presence of the ``missing mass'' problem in young regions such as Taurus (1-2 Myr). 
Alternatively, disk dust masses may be underestimated when using the millimeter flux method, which assumes that the disk is optically thin, whereas disks are unlikely to be optically thin throughout all millimeter wavelengths \citep{2019BalleringEisner, 2020Ribas,2023Xin,2023Rilinger}. Significant mass may also be ``hidden'' in optically thick dust traps \citep{2024Savvidou}.
Lastly, protoplanetary disks may replenish their material by accreting additional material from their surroundings \citep[e.g.][]{2008ThroopBally,2017Kuffmeier,2021Ginski}. 

\floattable
\begin{deluxetable}{l c c c c c c c c}
\tablecaption{Stellar parameters of the Serpens sample. \label{table:parameters_serpens}}
\tablehead{
\colhead{Object} &
\colhead{Parallax} &
\colhead{SpT} &
\colhead{$T_{eff}$} &
\colhead{$A_v$} &
\colhead{$L_*$} &
\colhead{$R_*$} &
\colhead{$M_*$} &
\colhead{Reference}
\\
\colhead{2MASS} & 
\colhead{mas} & 
\colhead{} & 
\colhead{K} & 
\colhead{} & 
\colhead{$L_{\odot}$} & 
\colhead{$R_{\odot}$} & 
\colhead{$M_{\odot}$} &
\colhead{}
}
\startdata
J18264598-0344306 & $2.21 \pm 0.21$ & M1 & $3730 \pm 100$ & $5.54 \pm 0.13$ & $0.04 \pm 0.03$ & $0.50 \pm 0.23$ & $0.50 \pm 0.28$ & 4\\
J18264973-0355596 & $2.26 \pm 0.16$ & K4 & $4800 \pm 100$ & $8.53 \pm 0.03$ & $0.20 \pm 0.17$ & $0.7 \pm 0.3$ & $0.7 \pm 0.4$ & 4\\
J18273709-0349385 & $2.63 \pm 0.11$ & K7 & $4234 \pm 100$ & $9.972 \pm 0.028$ & $0.10 \pm 0.08$ & $0.63 \pm 0.29$ & $0.6 \pm 0.4$ & 4\\
J18274426-0329519 & $2.48 \pm 0.17$ & M0 & $3890 \pm 100$ & $7.85 \pm 0.10$ & $0.07 \pm 0.06$ & $0.59 \pm 0.27$ & $0.6 \pm 0.3$ & 4\\
J18274445-0356016 & $2.29 \pm 0.07$ & F9.5 & $6008 \pm 100$ & $7.890 \pm 0.019$ & $4.21 \pm 0.22$ & $1.89 \pm 0.05$ & $1.22 \pm 0.04$ & 4\\
J18275383-0002335 & $2.37 \pm 0.04$ & K2 & $4900 \pm 500$ & $1.000 \pm 0.019$ & $2.0 \pm 1.8$ & $2.0 \pm 1.0$ & $1.6 \pm 0.4$ & 1\\
J18281519-0001408 & $2.58 \pm 0.22$ & M8 & $2640 \pm 330$ & $2.00 \pm 0.04$ & $0.11 \pm 0.07$ & $1.6 \pm 0.6$ & $0.12 \pm 0.12$ & 2\\
J18290088+0029312 & $2.37 \pm 0.04$ & K7 & $4060 \pm 220$ & $5.00 \pm 0.09$ & $1.0 \pm 0.9$ & $2.0 \pm 0.9$ & $1.14 \pm 0.20$ & 2\\
J18290394+0020212 & $2.322 \pm 0.027$ & K5 & $4400 \pm 300$ & $7.00 \pm 0.13$ & $5.0 \pm 3.0$ & $4.0 \pm 1.5$ & $1.6 \pm 0.4$ & 2\\
J18290935-0203503 & $2.02 \pm 0.16$ & K6 & $4360 \pm 100$ & $9.987 \pm 0.020$ & $0.14 \pm 0.11$ & $0.7 \pm 0.3$ & $0.7 \pm 0.4$ & 4\\
J18291969+0018030 & $2.23 \pm 0.21$ & M0 & $3850 \pm 100$ & $6.00 \pm 0.11$ & $0.6 \pm 0.4$ & $1.7 \pm 0.6$ & $0.90 \pm 0.10$ & 2\\
J18293084+0101071 & $2.37 \pm 0.06$ & F7 & $6290 \pm 120$ & $6.17 \pm 0.05$ & $2.5 \pm 2.0$ & $1.3 \pm 0.6$ & $1.2 \pm 0.7$ & 4\\
J18294147+0107378 & $2.33 \pm 0.04$ & K5 & $4400 \pm 300$ & $3.30 \pm 0.06$ & $0.9 \pm 0.4$ & $1.6 \pm 0.4$ & $0.7 \pm 0.3$ & 3\\
J18295360+0117017 & $1.99 \pm 0.10$ & K6 & $4200 \pm 200$ & $4.60 \pm 0.08$ & $0.52 \pm 0.22$ & $1.4 \pm 0.3$ & $0.7 \pm 0.3$ & 3\\
J18295820+0115216 & $2.43 \pm 0.14$ & K5.5 & $4300 \pm 200$ & $5.50 \pm 0.10$ & $0.31 \pm 0.12$ & $1.02 \pm 0.20$ & $0.8 \pm 0.4$ & 3\\
J18300545+0101022 & $2.39 \pm 0.22$ & M3 & $3440 \pm 100$ & $4.41 \pm 0.09$ & $0.016 \pm 0.013$ & $0.36 \pm 0.17$ & $0.37 \pm 0.20$ & 4\\
J18300610+0106170 & $2.37 \pm 0.08$ & M0 & $3800 \pm 210$ & $3.70 \pm 0.07$ & $0.51 \pm 0.21$ & $1.6 \pm 0.3$ & $0.42 \pm 0.20$ & 3\\
J18302240+0120439 & $2.33 \pm 0.05$ & M1 & $3630 \pm 130$ & $1.80 \pm 0.03$ & $0.26 \pm 0.12$ & $1.3 \pm 0.3$ & $0.41 \pm 0.19$ & 3\\
J18314416-0216182 & $1.99 \pm 0.10$ & F5 & $6650 \pm 100$ & $9.630 \pm 0.027$ & $3.6 \pm 0.9$ & $1.47 \pm 0.11$ & $1.33 \pm 0.08$ & 4\\
J18321676-0223167 & $1.91 \pm 0.16$ & K6 & $4409 \pm 100$ & $7.27 \pm 0.07$ & $0.14 \pm 0.11$ & $0.7 \pm 0.3$ & $0.7 \pm 0.4$ & 4\\
J18321707-0229531 & $2.12 \pm 0.22$ & K3 & $4850 \pm 100$ & $9.13 \pm 0.05$ & $0.28 \pm 0.23$ & $0.8 \pm 0.4$ & $0.8 \pm 0.4$ & 4\\
J18370677+0011479 & $1.89 \pm 0.21$ & M1 & $3690 \pm 100$ & $5.30 \pm 0.06$ & $0.04 \pm 0.03$ & $0.50 \pm 0.23$ & $0.50 \pm 0.28$ & 4\\
J18382267+0023093 & $2.22 \pm 0.20$ & M4.5 & $3187 \pm 100$ & $2.06 \pm 0.24$ & $0.004 \pm 0.003$ & $0.13 \pm 0.06$ & $0.18 \pm 0.1$ & 4
\enddata
\tablenotetext{*}{All parallax values are from \textit{Gaia} DR3. $R_*$ is calculated here using literature values. Additional parameters are obtained from various sources as follows: (1,2,3) spectral type, $T_{eff}$, $A_v$, $L_*$, and $M_*$ are compiled from \citet{2009Oliveira}, \citet{2013Oliveira}, and \citet{2015Erickson}, respectively; (4) spectral type is from \cite{2013PecautMamajek} using the $T_{eff}$ available in the literature; $T_{eff}$ and $A_v$ are gathered from the derived values from GSP-Phot using BP/RP spectra \citep{GaiaDR3}. The General Stellar Parametrizer for Photometry (GSP-Phot) derives stellar parameters ($T_{eff}$, $\log g$, $A_G$, distance, etc.) from BP/RP spectra down to G=19, assuming all sources are stars. It infers these parameters using Gaia’s low-resolution BP/RP spectra, astrometry, and photometry. We refer the reader to consult \cite{GaiaDR3} and \cite{2023Andrae} for further details on the GSP-Phot methodology. $L_*$ and $M_*$ are estimated here using the MassAge code following \cite{2023Hernandez}. In short, $L_*$ is calculated using extinction-corrected J magnitudes \citep{2020LuhmanEsplin}, bolometric corrections from \cite{2013PecautMamajek}, and distances derived from parallaxes (\textit{Gaia} DR3). $M_*$ is derived using the MIST \citep{2016Dotter} evolutionary models.}
\end{deluxetable}

Here, we use SED modeling to measure dust masses for a sample of disks in the Serpens and L1641/L1647 star-forming regions. We focus on disks around T-Tauri stars (TTS) which are pre-main sequence, low-mass \citep[0.08 M$_{\odot}$ $\leq M_* \leq$ 2M$_{\odot}$;][]{1945Joy, 1962Herbig}, cool (late spectral types K-M) stars exhibiting photometric variability as well as characteristic emission and absorption lines (e.g., H$\alpha$ and Li 6707 \AA, respectively). 
We use ALMA 1.3 mm continuum observations from the Serpens star-forming region \citep{2022Anderson} and the Orion A cloud in the L1641 and L1647 regions \citep{2022vanTerwisga}. We incorporate broad-band photometry and low-resolution \textit{Spitzer} Infrared Spectrograph (IRS) spectra from CASSIS \citep{2004IRS}. The SED model fitting adopts the technique developed by \cite{2020Ribas}, which employs an Artificial Neural Network (ANN) trained with the D'Alessio Irradiated Accretion Disk models \citep[DIAD;][]{1998Dalessio, 1999Dalessio, 2001Dalessio, 2005Dalessio, 2006Dalessio}. 
This paper's structure is as follows: we describe the sample selection in Section \ref{sec:sample}, we present analysis and results from the SED modeling in Section \ref{sec:analysis and results}, we discuss our results in Section \ref{sec:discussion}, and finally we provide conclusions in Section \ref{sec:summary}.

\floattable{\begin{deluxetable}{l c c c c c c c c}
\tablecaption{Stellar parameters for L1641/L1647.
\label{table:parameters_SODA}}
\tablehead{
\colhead{Object} &
\colhead{Parallax} &
\colhead{SpT} &
\colhead{$T_{eff}$} &
\colhead{$A_v$} &
\colhead{$L_*$} &
\colhead{$R_*$} &
\colhead{$M_*$} &
\colhead{Reference}
\\
\colhead{2MASS} & 
\colhead{mas} & 
\colhead{} & 
\colhead{K} & 
\colhead{} & 
\colhead{$L_{\odot}$} & 
\colhead{$R_{\odot}$} & 
\colhead{$M_{\odot}$} &
\colhead{}
}
\startdata
J05324993-0610456 & $2.63 \pm 0.06$	&	M0.8	&	$3770 \pm 110$	&	$0.90 \pm 0.24$	&	$0.08 \pm 0.23$	&	--	&	$0.55 \pm 0.07$	& 3,4\\
J05332098-0622333 & $2.54 \pm 0.05$	&	M1	&	$3742 \pm 100$	&	$2.1 \pm 0.6$	&	$0.4 \pm 1.1$	&	$1.6 \pm 2.2$	&	$1.80 \pm 0.22$	& 3,5,8\\
J05334160-0606074 & $2.44 \pm 0.15$	&	M3.5	&	$3378 \pm 100$	&	$0.17 \pm 0.05$	&	--	&	--	&	--	& 3,6\\
J05335945-0621182 & $2.65 \pm 0.07$	&	M4.1	&	$3210 \pm 100$	&	$0.30 \pm 0.08$	&	$0.09 \pm 0.25$	&	--	&	$0.170 \pm 0.021$	& 3,4\\
J05340108-0602268 & $2.600 \pm 0.018$	&	K7.6	&	$3860 \pm 100$	&	$0.34 \pm 0.09$	&	$0.4 \pm 1.0$	&	$1.4 \pm 1.9$	&	$0.68 \pm 0.04$	& 1,7\\
J05340495-0623469 & $2.61 \pm 0.05$	&	M4	&	$3349 \pm 100$	&	$0.94 \pm 0.21$	&	$0.2 \pm 1.1$	&	$1 \pm 4$	&	$0.35 \pm 0.04$	& 1,7\\
J05342043-0631150 & $2.42 \pm 0.15$	&	M5.1	&	$3060 \pm 100$	&	$0.60 \pm 0.16$	&	$0.06 \pm 0.16$	&	--	&	$0.1 \pm 0.012$	& 1,3,4\\
J05343096-0558036 & $2.32 \pm 0.09$	&	M5.1	&	$3100 \pm 100$	&	$0 \pm 0.3$	&	$0.05 \pm 0.14$	&	--	&	$0.098 \pm 0.012$	& 3,4\\
J05343319-0614287 & $2.64 \pm 0.20$	&	M5.4	&	$2930 \pm 100$	&	--	&	--	&	--	&	--	& 1,3\\
J05343360-0611194 & $2.58 \pm 0.07$ & M4.8 & $3110 \pm 100$ & $0 \pm 0.3$ & $0.06 \pm 0.17$ & -- & $0.120 \pm 0.015$ & 3,4
\enddata
\tablenotetext{*}{All parallax values are from \textit{Gaia} DR3. $R_*$ is calculated here using literature values. Additional parameters are obtained from various sources as follows: (1,2) spectral types come from \cite{2012Hsu} and \cite{2013Hsu}, respectively; (3) spectral type (or $T_{eff}$) is from \cite{2013PecautMamajek} using $T_{eff}$ (or spectral type) from the literature; (4) spectral type or $A_v$, $L_*$, and $M_*$ are sourced from \cite{2009Fang,2013FangA,2018Fang}; (5) $T_{eff}$ was taken from APOGEE-2 DR17 \citep{2023Abdurro}; (6) $T_{eff}$ and $A_v$ are gathered from the derived values from GSP-Phot using BP/RP spectra \citep{GaiaDR3} (see Table \ref{table:parameters_serpens} caption for further details on GSP-Phot); (7) \cite{2023Serna} provides $T_{eff}$, $A_v$, $L_*$, and $M_*$; (8) $A_v$, $L_*$, and $M_*$ is from \cite{2023Zuniga};
(9) \cite{2016Kim} gives spectral type or $A_v$, $L_*$, and $M_*$;  (10) \cite{2016DaRio} provides $A_v$, $L_*$, and $M_*$; (11) $T_{eff}$ or $L_*$, $R_*$ are gathered from Gaia DR2 \citep{2018GaiaDR2} estimates for temperature from Apsis-Priam and radius/luminosity estimates from Apsis-FLAME, respectively (see \cite{2018Andrae} and \cite{2013BailerJones} for more details regarding the stellar parameter estimation with the Apsis data processing pipeline); (12) stellar parameters correspond to those adopted for the objects analyzed in \cite{2023Rilinger}.
\\
This table is published in its entirety in the machine-readable format. A portion is shown here for guidance regarding its form and content.}
\end{deluxetable}}

\section{Sample selection} \label{sec:sample}
The sample used in this work comprises TTS with disks in the Serpens, L1641, and L1647 regions. (Sub)millimeter observations are necessary to constrain the disk's dust mass, so we compiled our sample using (sub)millimeter surveys of these regions. We selected our Serpens sample from \cite{2022Anderson}, a continuum and $^{12}CO\ (J=2-1)$ line survey spread over an area of $\sim$10 square degrees in Serpens which observed 320 objects studied by the \textit{Spitzer} From Molecular Cores to Planet-Forming Disks C2D Overview \citep[hereafter c2d; ][]{SpitzerEvans, 2009Evans} and the Gould Belt Survey \citep{2015Dunham}. The L1641 and L1647 sample is obtained from the Survey of Orion Disks with ALMA \citep[SODA;][]{2022vanTerwisga}, which observed 873 protoplanetary disks identified by \textit{Spitzer} \citep{2009Evans, 2015Dunham}.  Both \cite{2022Anderson} $\&$ \cite{2022vanTerwisga} use the ALMA 1.3 mm band (centered at 241.1 GHz and 225 GHz, respectively). 

The original sample of 320 Serpens sources \citep{2022Anderson} comes down to 23 objects, and the 873 L1641/L1647 sources \citep{2022vanTerwisga} to 188 objects after considering several criteria. They must belong to the Serpens or L1641/L1647 star-forming regions and be identified as Class II YSOs. Each source must have been detected by ALMA and have at least two 2MASS measurements, two mid-infrared (\textit{WISE} or c2d) measurements, and one (sub)millimeter (ALMA) measurement. Sources known to be transitional disks (TDs), exhibit flat spectra, or belong to binary or multiple systems were excluded. Additionally, the sources must not have a (sub)mm measurement within 2''.5 of another (sub)mm detection of at least 3$\sigma$. Only sources with ALMA detections of at least 3$\sigma$ were included. To ensure reliable astrometric data (e.g., parallax), the selected sources must display good \textit{Gaia} astrometric quality. Finally, all sources must be classified as T Tauri stars and have stellar parameters available in the literature. The $T_{eff}$, $R_{*}$, $M_{*}$, and parallaxes listed in Tables \ref{table:parameters_serpens} and \ref{table:parameters_SODA} are used in the SED fitting as described in Section \ref{subsec:sed modeling ANN} and \ref{subsubsec:sed modeling MCMC}, and the rest of the parameters are listed for completeness. 
We note that for $T_{eff}$, we adopted an uncertainty of 100 K for those objects with reported uncertainties below 100 K, following \cite{2020Ribas}.

We refer the reader to Appendix \ref{sec:appendixA} for a more detailed review of the selection criteria. This results in an overall sample of 211 objects.

\section{Analysis and Results} 
\label{sec:analysis and results}
Below, we compile the SEDs for the 211 sources in Serpens, L1641, and L1647 selected in Section \ref{sec:sample}. A subset of 41 sources from L1641 were modeled in \citet{2023Rilinger} and we incorporate their model results here. We provide an overview of the SED modeling methodology using the ANN developed by \citet{2020Ribas}, which applies a Bayesian approach and is trained with DIAD models. We end with the resulting model fits and further reduce our sample to 178 sources (18 from Serpens and 160 from L1641/L1647) based on the quality of the fits.

\subsection{SEDs} 
\label{subsubsec: SED modeling intro}
We compile the SEDs using multiwavelength photometry gathered for each object from the Vizier catalog access tool \citep{2000Vizier}, using a search radius of 2''. Table \ref{table:photometry} lists all photometry sources used. We also include low-resolution \textit{Spitzer} IRS spectra for each object, if available, which were obtained from the Combined Atlas of Sources with \textit{Spitzer} IRS Spectra \citep[CASSIS;][]{2011CASSIS}.

\floattable{\begin{deluxetable}{l c l r}
\tablecaption{Photometry sources for the sample \label{table:photometry}}
\tablehead{
\colhead{SED filter} &
\colhead{Wavelength [$\mu m$]} &
\colhead{Instrument} & \colhead{Reference}}
\startdata
SDSS:u & 0.352 & APOGEE-2 & \cite{SDSS}\\
Johnson B & 0.4361 & AAVSO & \cite{2016Henden}\\
POSS-II:J & 0.468 & GSC2.3 & \cite{GSC2.3Lasker}\\
PS1:g & 0.477 & PAN-STARRS & \cite{PANSTARRS}\\
SDSS:g & 0.482 & APOGEE-2 & \cite{SDSS}\\
G$_{BP}$ & 0.504 & \textit{Gaia} DR3 & \cite{GaiaDR3}\\
Johnson V & 0.5448 & AAVSO & \cite{2016Henden}\\
G & 0.582 & \textit{Gaia} DR3 & \cite{GaiaDR3}\\
PS1:r & 0.613 & PAN-STARRS & \cite{PANSTARRS}\\
SDSS:r & 0.625 & APOGEE-2 & \cite{SDSS}\\
POSS-II:F & 0.640 & GSC2.3 & \cite{GSC2.3Lasker}\\
G$_{RP}$ & 0.726 & \textit{Gaia} DR3 & \cite{GaiaDR3}\\
PS1:i & 0.748 & PAN-STARRS & \cite{PANSTARRS}\\
SDSS:i & 0.763 & APOGEE-2 & \cite{SDSS}\\
POSS-II:i & 0.784 & GSC2.3 & \cite{GSC2.3Lasker}\\
PS1:z & 0.865 & PAN-STARRS & \cite{PANSTARRS}\\
SDSS:z & 0.902 & APOGEE-2 & \cite{SDSS}\\
PS1:y & 0.960 & PAN-STARRS & \cite{PANSTARRS}\\
W3 & 1.16 & \textit{WISE} & \cite{ALLWISECutri}\\
J & 1.24 & 2MASS & \cite{2MASSSkrutskie}\\
H & 1.65 & 2MASS & \cite{2MASSSkrutskie}\\
K$_{s}$ & 2.16 & 2MASS & \cite{2MASSSkrutskie}\\
W4 & 2.21 & \textit{WISE} & \cite{ALLWISECutri}\\
W1 & 3.35 & \textit{WISE} & \cite{ALLWISECutri}\\
\textit{IRAC}:3.6 & 3.55 & \textit{Spitzer} & \cite{SpitzerEvans}\\
\textit{IRAC}:4.5 & 4.49 & \textit{Spitzer} & \cite{SpitzerEvans}\\
W2 & 4.6 & \textit{WISE} & \cite{ALLWISECutri}\\
\textit{IRAC}:5.8 & 5.73 & \textit{Spitzer} & \cite{SpitzerEvans}\\
\textit{IRAC}:8.0 & 7.87 & \textit{Spitzer} & \cite{SpitzerEvans}\\
\textit{MIPS}:24 & 23.7 & \textit{Spitzer} & \cite{SpitzerEvans}\\
\textit{Herschel}:70 & 70 & \textit{Herschel} & \cite{2017Herschel}\\
\textit{Herschel}:100 & 100 & \textit{Herschel} & \cite{2017Herschel}\\
\textit{Herschel}:160 & 160 & \textit{Herschel} & \cite{2017Herschel}\\
Band 6 & 1333 & ALMA & \cite{2022Anderson, 2022vanTerwisga}
\enddata
\end{deluxetable}}

\subsection{SED modeling} 
To fit our objects' SEDs, we use a model that relies on the ANN developed by \cite{2020Ribas}, and trained on the D’Alessio Irradiated Accretion Disk models \citep[DIAD;][]{1998Dalessio,1999Dalessio,2001Dalessio,2005Dalessio,2006Dalessio}. We follow a Bayesian approach as detailed in \cite{2020Ribas} and \cite{2023Xin} to estimate our sample of disks' posterior probability density functions based on their respective SEDs.

\subsubsection{The model} \label{subsec:sed modeling ANN}
 The work by \cite{2020Ribas} presents a novel approach to fit samples of protoplanetary disks, combining physically motivated models with a neural network to speed the process of model fitting; the ANN takes a few milliseconds to evaluate the SEDs, compared to the 1-2 hours that the DIAD models require, which, as previously discussed, makes the model fitting considerably time-consuming for large samples. We direct the reader to Section 4 and Figure 2 of \cite{2020Ribas} for a detailed description of the model and a summary of the steps, which are also summarized below.

The ANN comprises two artificial neural networks: ANN$_{SED}$ for SED estimates and ANN$_{diskmass}$ for disk mass estimates. The ANN provides SED estimates based on input parameters. 
Each of the models computed in the fitting procedure goes through the following steps:
\begin{enumerate}
    \item Stellar age and $M_*$ are taken as free parameters of the model, and for each pair, $T_*$ and $R_*$ are estimated from the MESA Isochrones and Stellar Tracks \citep[MIST;][]{2011Paxton, 2013Paxton, 2015Paxton, 2016Dotter, 2016Choi}. This process ensures that our model explores combinations of stellar age, $M_*$, $R_*$, and $T_*$ that are physically realistic, which is required because these parameters are used to evaluate the SEDs. Likewise, $L_*$ is not a free parameter in the model but is derived from MIST. We point the reader to Section \ref{subsubsec:sed modeling MCMC} for more details about how literature information about the stellar parameters is used in the modeling.
    \item The ANN uses $M_*$, $T_*$, and $R_*$ as well as the following disk and dust parameters:
    \begin{itemize}
        \item Mass accretion rate $\dot{M}$
        \item Disk viscosity $\alpha$ \citep[adimensional and following][]{1973Shakura}.
        \item Dust settling $\epsilon$ (adimensional)
        \item Maximum grain sizes in the disk atmosphere $a_{max,upper}$ and midplane $a_{max,midplane}$
        \item Disk size $R_{disk}$
        \item Inclination $i$ 
        \item Dust sublimation temperature $T_{wall}$
        \item Scaling factor of the inner wall $z_{wall}$
    \end{itemize}
    Consequently, the ANN generates the corresponding SED for a reference distance of 100 pc in the absence of extinction to the source. We refer the reader to the DIAD papers \citep{1998Dalessio,1999Dalessio,2001Dalessio,2005Dalessio,2006Dalessio} for further details on these disk and dust parameters.
    \item The SED output is scaled to the source's distance (based on the corresponding parallax, which is another free parameter in our model) and reddened by $A_v$, applying the \citet{2009McClure} extinction law \citep[following][]{2017Ribas}. Here, $A_v$ denotes the interstellar extinction and does not include the self-extinction contributed by the disk, which is already incorporated in DIAD.
\end{enumerate}

Some caveats need to be taken into account for this methodology. As discussed in Section \ref{sec:sample}, the objects that make up our sample must comply with specific selection criteria.
Some physical properties remain fixed (e.g. dust composition, gas-to-dust ratio) or not considered in the modeling (i.e. the presence of substructures in the disk). Refer to \citet{2020Ribas} for an in-depth discussion of the caveats of this methodology. Most importantly, this sample is only made up of sources detected in the (sub)mm range. This would bias the analysis towards massive disks and underrepresented  low-mass disks, which are less luminous, smaller in radius, and thus harder to detect. 

\subsubsection{Markov Chain Monte Carlo (MCMC)}
\label{subsubsec:sed modeling MCMC}
We implement a Bayesian analysis framework as a Markov Chain Monte Carlo (MCMC) fitting process to probe the large parameter space and estimate posterior distributions for the model parameters. Following \citet{2023Rilinger}, we adopted the likelihood functions from \citet{2020Ribas} and \citet{2023Xin}, as well as the \texttt{ptemcee} \citep{2016Vousden} parallel-tempering version of the \texttt{emcee} MCMC code \citep{2013Foreman-Mackey}. Our photometry points were divided into two groups: 
\begin{itemize}
    \item critical: includes 2MASS and ALMA photometry \citep{2022Anderson}
    \item general: includes all other photometry
\end{itemize}
The division allows us to provide further weight to the 2MASS and ALMA photometry, which are vital for accurately scaling the stellar photosphere and calculating the disk mass, respectively. Four standard Gaussian likelihood functions are used, each for the critical photometry data, the spectral data (i.e., IRS Spectra), the $T_*$, and the $R_*$. To account for possible systematic uncertainties that may be present in the data from different instruments, \cite{2023Xin} scaled all of the uncertainties in the flux densities according to the free parameter, \textit{f}. Some photometric points may differ significantly from the rest of the SEDs and are problematic in the fitting process. \cite{2023Xin} employed a mixture model approach, following the methodology described by \cite{2010Hogg}. This approach was applied to the general photometry data, enabling the overall likelihood to incorporate information from both the photometric and spectral data. We note that Gaussian priors are used for the $M_{*}$ and parallaxes listed in Tables \ref{table:parameters_serpens} and \ref{table:parameters_SODA} to represent pre-existing information on these parameters, using their values as the centers and their uncertainties as the standard deviations, while flat priors are used when literature values are absent (e.g., inclination). We refer the reader to consult \cite{2023Xin} and \cite{2023Rilinger} for more details.

Hence, the model has 13 free parameters: Stellar age, $M_*$, DIAD parameters (i.e. $\dot{M}$, $\alpha$, $\epsilon$, $a_{max,upper}$, $a_{max,midplane}$, $R_{disk}$, $i$, $T_{wall}$, and  $z_{wall}$), parallax, and $A_v$ of the source, all of them determined using the MCMC fitting process. Finally, since we are compiling the SEDs from multiple catalogs, it is likely that some of the data points may be affected by undetected issues (i.e., source blending, saturation, artifacts). In order to reduce the impact of these outliers on the modeling, we include four additional parameters to mitigate the influence of such uncertainties during the fitting process: three of these account for possible outlier photometric points, and one, the \textit{f} parameter, accounts for potential systematic uncertainties inherent in the data acquired from different instruments. This results in a total of 17 free parameters, which are varied during the fitting to compute posterior probability distributions.

For an in-depth exploration and discussion of the MCMC process on which the ANN operates, we refer to the following works by \citet{2020Ribas}, \citet{2023Rilinger}, and \citet{2023Xin}.

\subsection{Model results} 
\label{subsec:model results}

\begin{figure*}
    \centering
    \includegraphics[width=17cm]{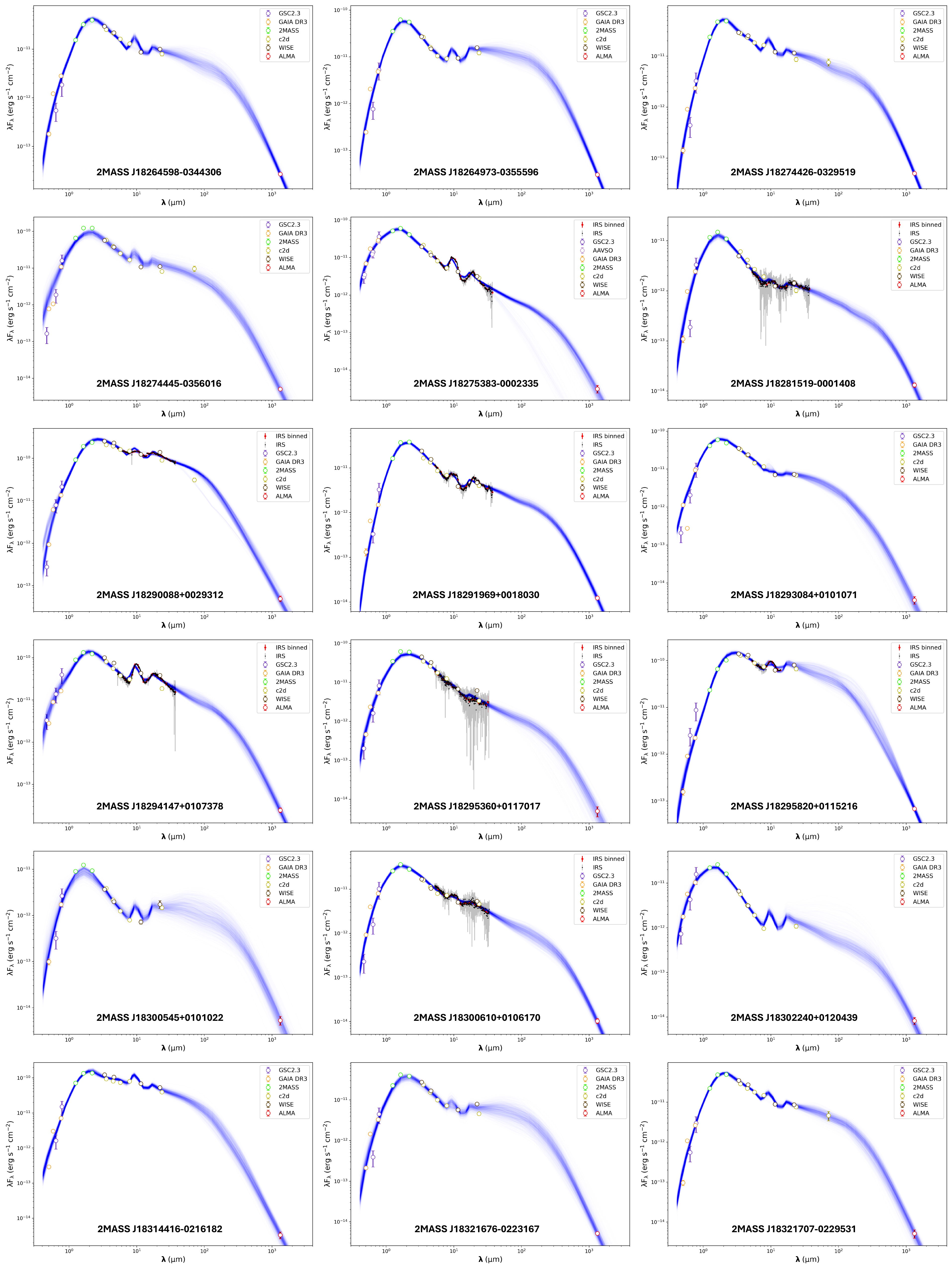}
    \caption{The observed SEDs and models for the 18 successfully fitted protoplanetary disks in our Serpens sample. The blue lines show 1000 randomly selected models from the posterior distributions. The models are reddened using \cite{2009McClure} extinction law and the extinction values gathered from the MCMC fitting process (see Section \ref{subsubsec:sed modeling MCMC}). Photometry labels are detailed in Table \ref{table:photometry}. Spitzer data is labeled as c2d. The complete figure set (18 images) is available in the online journal.}
    \label{fig:SED_models_serpens}
\end{figure*}

\begin{figure*}
    \centering
    \includegraphics[width=\textwidth]{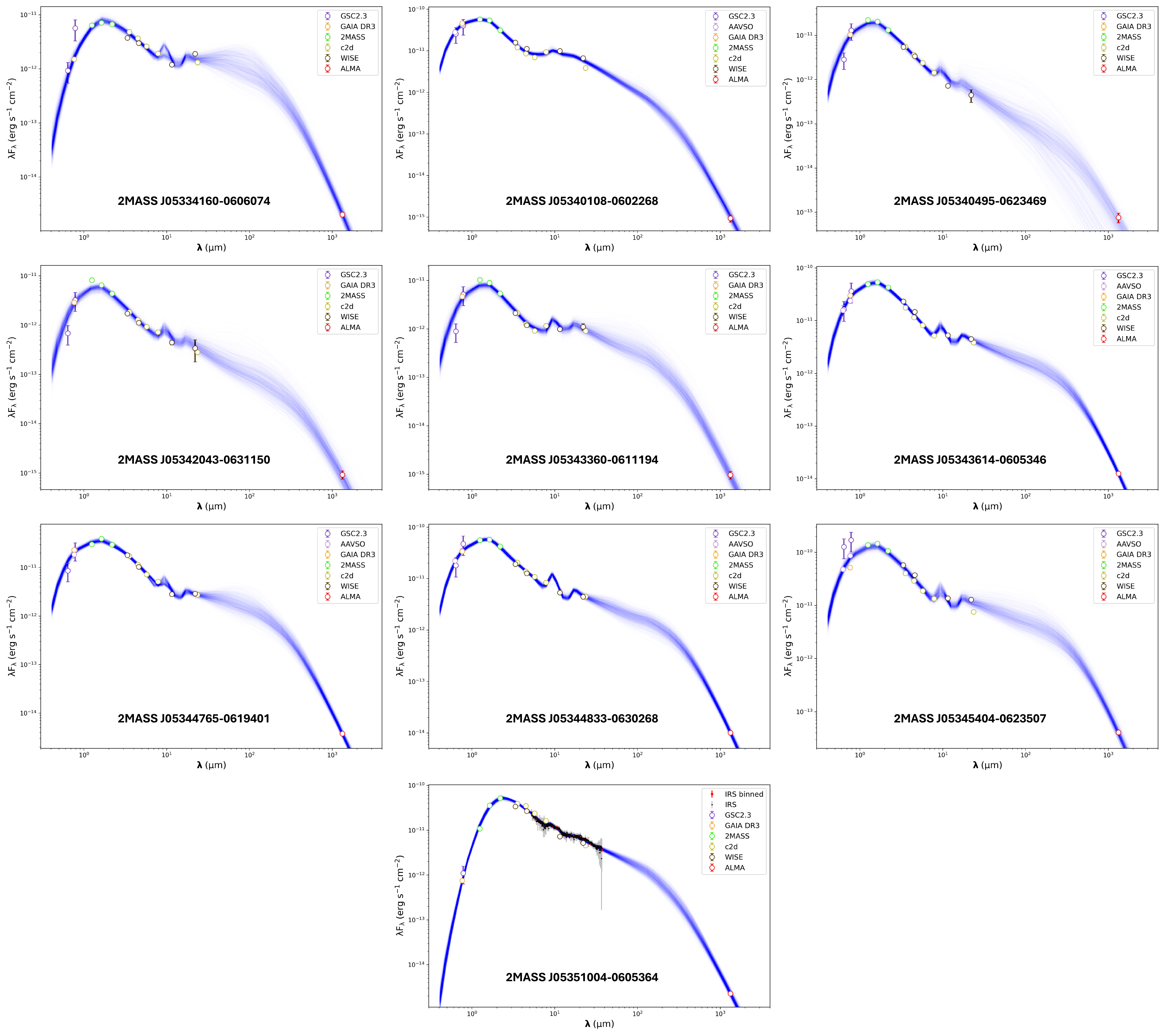}
    \caption{The observed SEDs and successfully fitted models for the first 10 of the 160 protoplanetary disks in our L1641/L1647 sample. The blue lines show 1000 randomly selected models from the posterior distributions. The models are reddened using \cite{2009McClure} extinction law and the extinction values gathered from the MCMC fitting process (see Section \ref{subsubsec:sed modeling MCMC}). Photometry labels are detailed in Table \ref{table:photometry}. Spitzer data is labeled as c2d. The complete figure set (119 images) is available in the online journal. The additional SEDs for the 41 L1641 objects can be consulted in \cite{2023Rilinger}.} 
    \label{fig: SED models_SODA}
\end{figure*}

Figures \ref{fig:SED_models_serpens} $\&$ \ref{fig: SED models_SODA} display the modeling results for Serpens and L1641/L1647, respectively. 
We identified successful and unsuccessful model fits for both samples; see \cite{2023Rilinger} for examples and further discussion on unsuccessful model fits.  

For the Serpens sample, six sources have unsuccessful fits, which may be due to incorrect stellar parameters, unknown substructures, or the potential presence of companions, as well as processes not taken into account by the methodology of \citet{2020Ribas}. These include the following sources:
2MASS J18273709-0349385 (model does not fit the \textit{Spitzer} photometry),
2MASS J18290394+0020212 (exhibits variability between the infrared IRS spectra and \textit{Spitzer} photometry),
2MASS J18290935-0203503 (model does not fit infrared photometry, \textit{Spitzer} and \textit{WISE}, as well as the millimeter photometry),
2MASS J18370677+0011479 (the model does not fit the 2MASS photometry and infrared structure of the SED),
2MASS J18382267+0023093 (the model does not fit the infrared structure of the SED), 2MASS J18373580+0019053 (the photometry has very large scatter and is difficult to fit).
Thus, the Serpens sample consists of 18 successfully modeled disks. 

For the L1641/L1647 region, 28 sources were discarded based on failed fits. We arrive at a sample of 160 successfully modeled sources in L1641/L1647.

Results from the model fits for individual sources and ensemble distributions are shown in Figures \ref{fig: individual_ensemble_serpens_distributions} and \ref{fig: individual_ensemble_SODA_distributions} for Serpens and L1641/L1647, respectively. Similarly, Table \ref{table: ANN_posteriors_Serpens_stellar} display the median values and corresponding errors (calculated from the 16th and 84th percentiles of the distribution) for the posterior distributions of stellar and disk parameters, respectively, obtained from the model fitting for Serpens. Meanwhile, Table \ref{tab: ANN_posteriors_SODA_stellar} lists those for L1641/L1647.
An assessment of the consistency between the stellar parameters ($M_*$, $R_*$, parallax, and $T_{\mathrm{eff}}$; as listed in Tables \ref{table:parameters_serpens} and \ref{table:parameters_SODA}, and their respective posterior distributions in Tables \ref{table: ANN_posteriors_Serpens_stellar} and \ref{tab: ANN_posteriors_SODA_stellar}) is conducted to validate the robustness of the ANN fitting methodology. We refer the reader to Appendix \ref{sec:appendixB} for a more detailed review of this validation procedure.
The corresponding corner plots generated from the MCMC process for all 178 disks are hosted within the designated \href{https://zenodo.org/records/17203297?token=eyJhbGciOiJIUzUxMiJ9.eyJpZCI6IjAwMTFmM2YzLTgxZDgtNGVhMy05NzUwLTA5MGZhZWZjNWQ3MCIsImRhdGEiOnt9LCJyYW5kb20iOiJlMmI1YTRlMGJiYzQ0NGExOGY5YTUwZWM4M2ZiYmQ1ZCJ9.mjCth0bpYmepMixe0paJy4MM67Z2mXDiXHJzJoYzOFLeTxjK16C2yLh80Mfe9qsVxD8yjVOy2Ann9UHTfLI73w}{Zenodo} repository.

Figures \ref{fig: individual_ensemble_serpens_distributions} \& \ref{fig: individual_ensemble_SODA_distributions} show some similarities in trends of the physical parameters. 
Disk mass depends on the disk viscosity ($\alpha$) and the mass accretion rate ($\dot{M}$), as both parameters determine the surface density profile of the disk $\Sigma \propto {\dot{M}}/{\alpha}$. These parameters are relatively well-constrained for both samples compared to the other parameters. The $R_{disk}$ and $a_{max,midplane}$ (i.e., the maximum grain size in the midplane) are not well constrained, consistent with the results of \cite{2020Ribas} and \cite{2023Xin}, which use the same method. 
The $\epsilon$ parameter, which measures the dust settling in the disk, shows a preference towards lower values in both samples (i.e., $\epsilon < 10^{-2}$), indicating that the studied disk populations have already undergone significant settling. This agrees with previous works where high degrees of dust settling are observed in L1641/L1647 \citep{2018Grant, 2019Grossschedl} and Serpens \citep{2023Liu}.  
For $a_{max,upper}$, the maximum grain size in the upper layers of the disk, some objects in both samples are well-constrained while others are not. This may be due to the availability of IRS spectra for each object, as the size of the upper layer dust grains is determined from mid-IR, in particular from the shape of the silicate feature at 10-20 $\mu$m. This reinforces the importance of having mid-IR spectra to better constrain dust population characteristics in the disk. 
Further discussion of a similar analysis of these posterior distributions for different populations can be found in \citet{2020Ribas} for Taurus-Auriga and in \citet{2023Xin} for Lupus.

\begin{figure*}
    \centering
    \includegraphics[width=17cm]{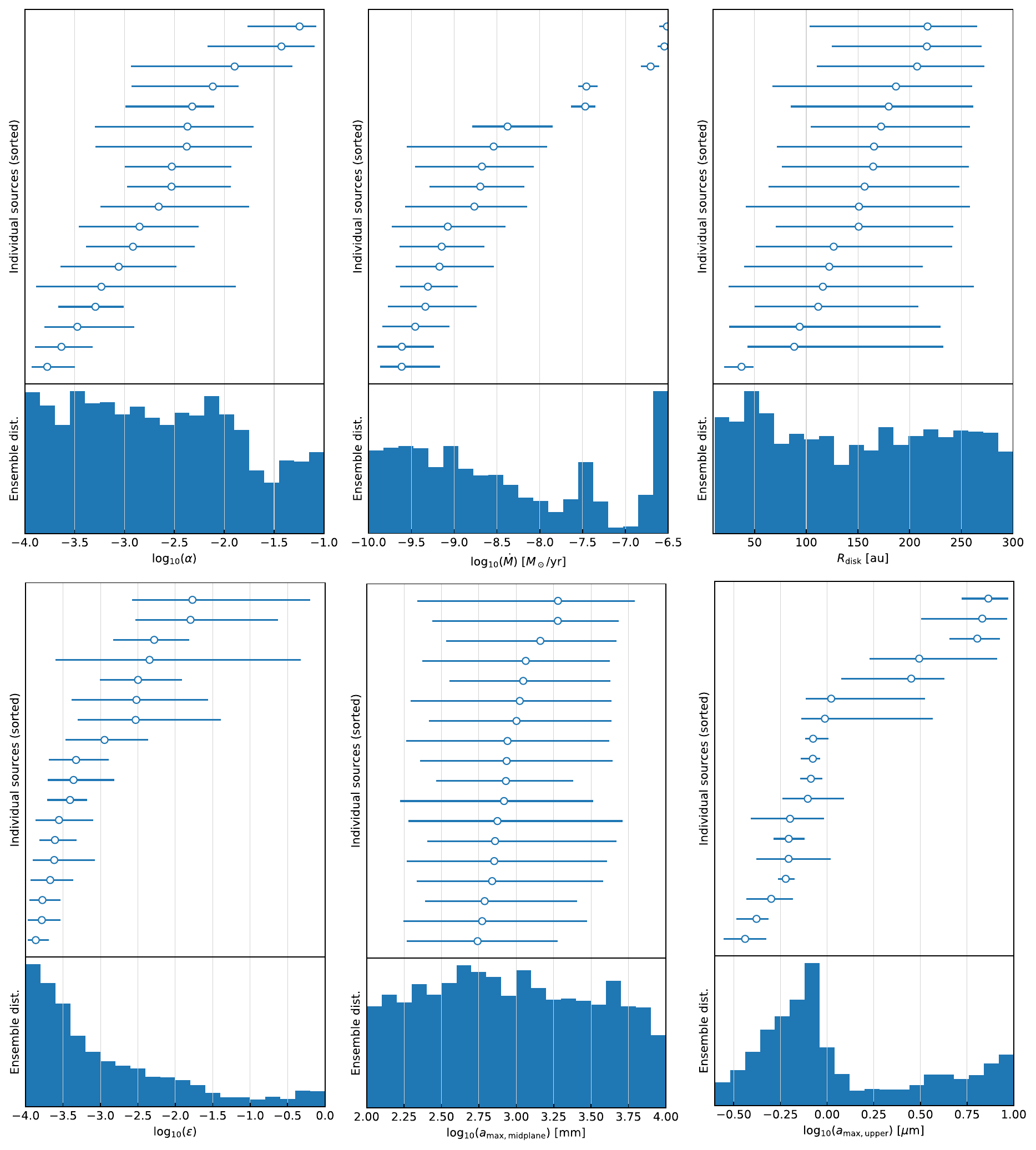}
    \caption{Posterior distributions of the individual 18 successfully modeled sources of the Serpens sample (top panels) and ensemble distribution (bottom panels) for the following parameters (from top left to bottom right): $\alpha$ (disk viscosity), $\dot{M}$ (mass accretion rate), $R_{disk}$ (outer disk radius), $\epsilon$ (dust settling), $a_{max,midplane}$ (maximum grain size in the midplane), $a_{max,upper}$ (maximum grain size in the upper layers of the disk). The error bars denote the 16th and 84th percentiles; the bottom panels' histograms use 20 bins.}
    \label{fig: individual_ensemble_serpens_distributions}
\end{figure*}

\begin{figure*}
    \figurenum{3}
    \centering
    \includegraphics[width=11cm]{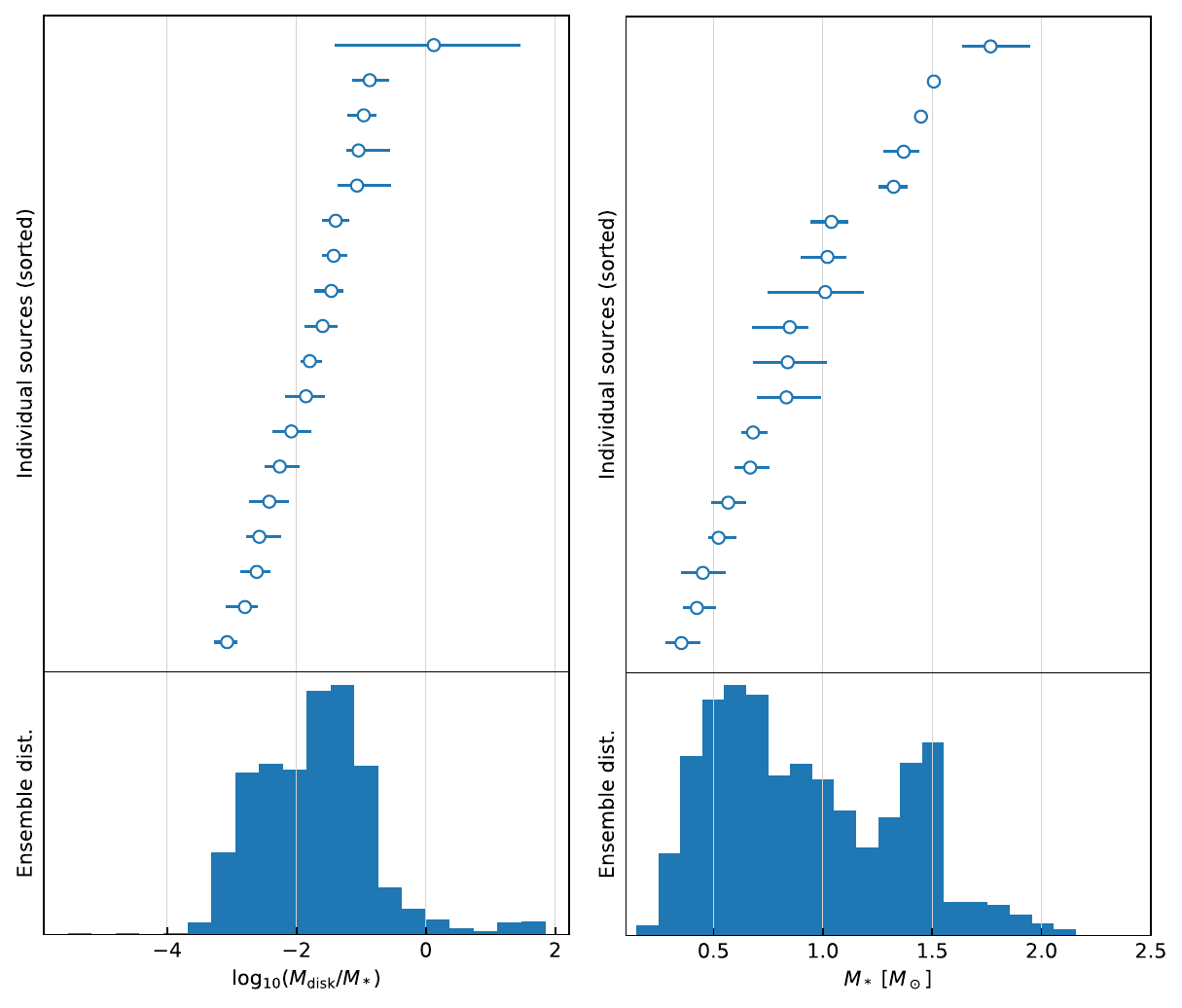}
    \caption{(Continued). Posterior distributions of the individual 18 successfully modeled sources of the Serpens sample (top panels) and ensemble distribution (bottom panels) for the following parameters (from left to right): $M_{disk}/M_*$ (disk mass fraction to host star mass), $M_*$ (host star mass).}
\end{figure*}

\begin{figure*}
    \centering
    \includegraphics[width=17cm]{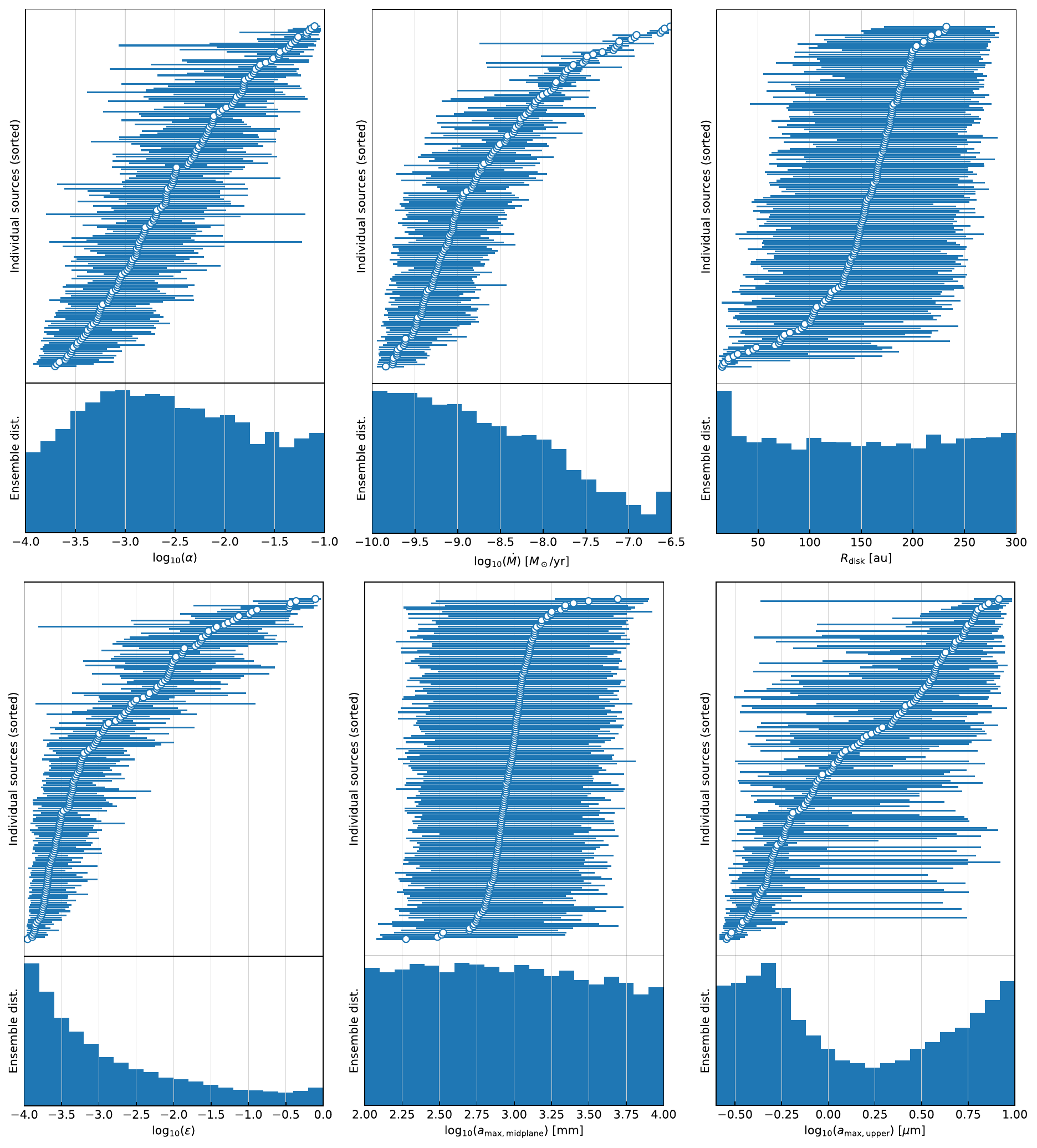}
    \caption{Posterior distributions of the individual 160 successfully modeled sources of the L1641/L1647 sample (top panels) and ensemble distribution (bottom panels) for the following parameters (from top left to bottom right): $\alpha$ (disk viscosity), $\dot{M}$ (mass accretion rate), $R_{disk}$ (outer disk radius), $\epsilon$ (dust settling), $a_{max,midplane}$ (maximum grain size in the midplane), $a_{max,upper}$ (maximum grain size in the upper layers of the disk). The error bars denote the 16th and 84th percentiles; the bottom panels' histograms use 20 bins.}
    \label{fig: individual_ensemble_SODA_distributions}
\end{figure*}

\begin{figure*}
    \figurenum{4}
    \centering
    \includegraphics[width=11cm]{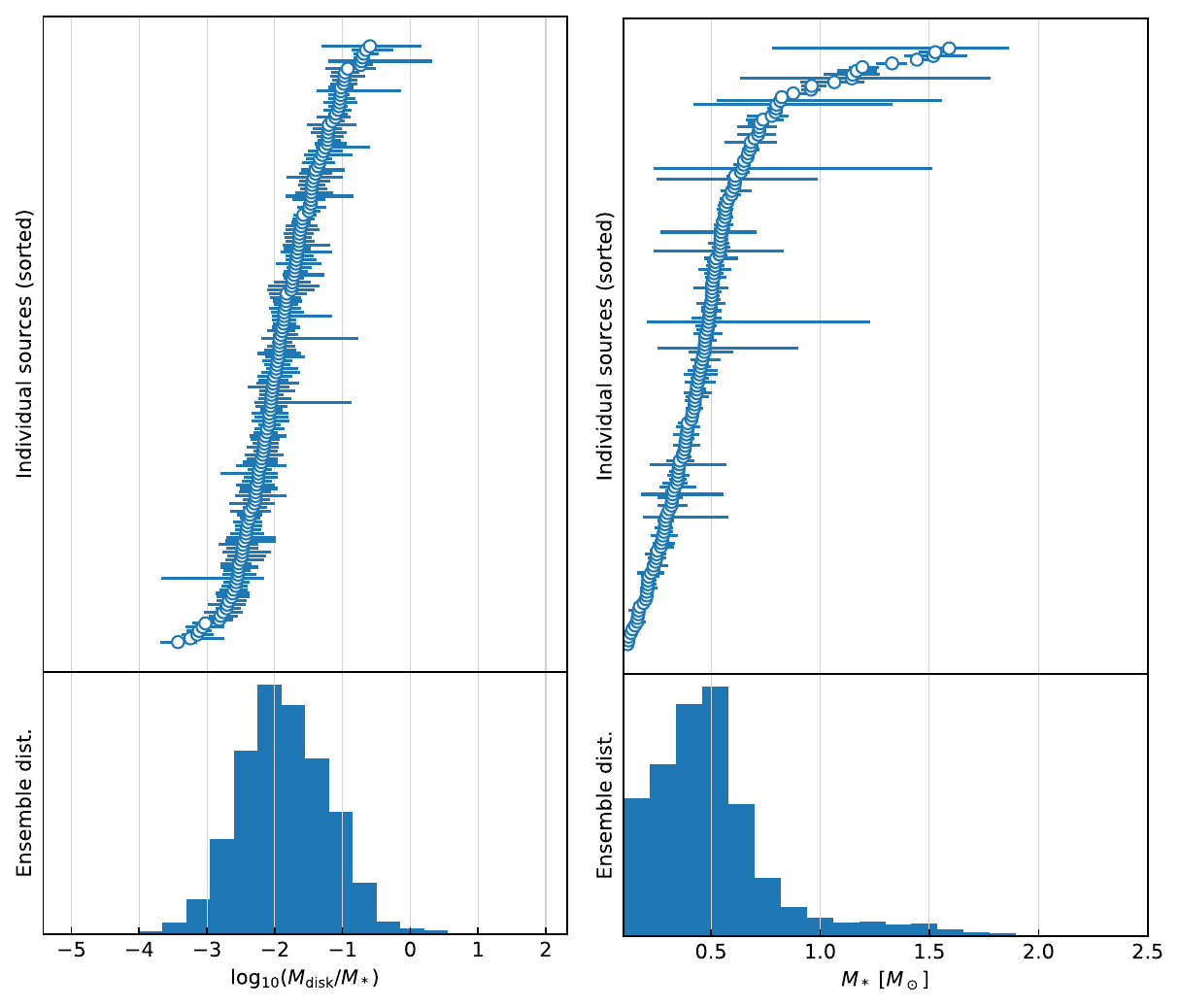}
    \caption{(Continued). Posterior distributions of the individual 160 successfully modeled sources of the L1641/L1647 sample (top panels) and ensemble distribution (bottom panels) for the following parameters (from left to right): $M_{disk}/M_*$ (disk mass fraction to host star mass), $M_*$ (host star mass).}
\end{figure*}

\floattable
\begin{splitdeluxetable*}{lcccccBcccccccc}
\tablecaption{Posteriors for Serpens stellar and disk parameters \label{table: ANN_posteriors_Serpens_stellar}}
\tablehead{
\colhead{Object} &
\colhead{$T_{eff}$} &
\colhead{Radius} &
\colhead{Mass} &
\colhead{Parallax} &
\colhead{$log_{10}(\dot{M})$} & 
\colhead{$log_{10}(\alpha)$} &
\colhead{$log_{10}(\epsilon)$} &
\colhead{$R_{disk}$} &
\colhead{$\mathbf{a_{\max,\;upper}}$} &
\colhead{$\mathbf{a_{\max,\;midplane}}$} &
\colhead{$T_{wall}$} &
\colhead{Inclination} &
\colhead{$M_{dust}$}
\\
\colhead{2MASS} & 
\colhead{K} & 
\colhead{$R_{\odot}$} & 
\colhead{$M_{\odot}$} & 
\colhead{mas} &
\colhead{$M_{\odot}/yr$} &
\colhead{} &
\colhead{} &
\colhead{au} &
\colhead{$\mu$m} &
\colhead{mm} &
\colhead{K} &
\colhead{deg} &
\colhead{$M_{\oplus}$}
}
\startdata
 J18264598-0344306 & $3720_{-90}^{+80}$ & $1.02_{-0.09}^{+0.14}$ & $0.54_{-0.07}^{+0.08}$ & $2.3_{-0.2}^{+0.2}$ & $-7.50_{-0.12}^{+0.13}$ & $-2.1_{-0.9}^{+0.3}$ & $-2.4_{-0.9}^{+1.3}$ & $100_{-60}^{+120}$ & $0.86_{-0.09}^{+0.21}$ & $1000_{-810}^{+4000}$ & $1400_{-40}^{+50}$ & $59_{-12}^{+6}$ & $200_{-90}^{+700}$ \\
 J18264973-0355596 & $4800_{-100}^{+100}$ & $1.93_{-0.16}^{+0.17}$ & $1.37_{-0.09}^{+0.08}$ & $2.48_{-0.14}^{+0.14}$ & $-8.7_{-0.7}^{+0.5}$ & $-3.1_{-0.6}^{+0.6}$ & $-2.9_{-0.5}^{+0.6}$ & $150_{-93}^{+100}$ & $0.51_{-0.14}^{+0.22}$ & $1000_{-700}^{+3000}$ & $1420_{-50}^{+50}$ & $20_{-10}^{+30}$ & $70_{-20}^{+50}$ \\
 J18274426-0329519 & $3880_{-90}^{+90}$ & $1.25_{-0.16}^{+0.21}$ & $0.66_{-0.08}^{+0.08}$ & $2.58_{-0.17}^{+0.17}$ & $-7.50_{-0.13}^{+0.12}$ & $-2.3_{-0.5}^{+0.2}$ & $-3.3_{-0.4}^{+0.5}$ & $169_{-100}^{+92}$ & $0.62_{-0.06}^{+0.07}$ & $1000_{-800}^{+3000}$ & $1440_{-40}^{+40}$ & $59_{-8}^{+6}$ & $300_{-90}^{+200}$ \\
 J18274445-0356016 & $5860_{-50}^{+60}$ & $1.85_{-0.04}^{+0.04}$ & $1.450_{-0.012}^{+0.013}$ & $2.31_{-0.07}^{+0.07}$ & $-9.1_{-0.6}^{+0.7}$ & $-2.6_{-0.6}^{+0.7}$ & $-3.6_{-0.3}^{+0.6}$ & $180_{-100}^{+80}$ & $0.76_{-0.18}^{+0.50}$ & $1000_{-500}^{+3000}$ & $1420_{-50}^{+50}$ & $22_{-9}^{+21}$ & $11_{-4}^{+8}$ \\
 J18275383-0002335 & $4400_{-300}^{+300}$ & $1.57_{-0.12}^{+0.15}$ & $1.03_{-0.27}^{+0.16}$ & $2.36_{-0.04}^{+0.04}$ & $-9.6_{-0.3}^{+0.5}$ & $-2.9_{-0.4}^{+0.6}$ & $-3.7_{-0.2}^{+0.3}$ & $180_{-100}^{+90}$ & $0.36_{-0.07}^{+0.10}$ & $1000_{-1100}^{+4000}$ & $1400_{-50}^{+50}$ & $68_{-4}^{+1}$ & $12_{-6}^{+10}$ \\
\enddata
\tablecomments{This table is published in its entirety in the machine-readable format. A portion is shown here for guidance regarding its form and content.}
\end{splitdeluxetable*}

\floattable
\begin{splitdeluxetable*}{lcccccBcccccccc}
\tablecaption{Posteriors for L1641/L1647 stellar and disk parameters \label{tab: ANN_posteriors_SODA_stellar}}
\tablehead{
\colhead{Object} &
\colhead{$T_{eff}$} &
\colhead{Radius} &
\colhead{Mass} &
\colhead{Parallax} &
\colhead{$log_{10}(\dot{M})$} & 
\colhead{$log_{10}(\alpha)$} &
\colhead{$log_{10}(\epsilon)$} &
\colhead{$R_{disk}$} &
\colhead{$\mathbf{a_{\max,\;upper}}$} &
\colhead{$\mathbf{a_{\max,\;midplane}}$} &
\colhead{$T_{wall}$} &
\colhead{Inclination} &
\colhead{$M_{dust}$}
\\
\colhead{2MASS} & 
\colhead{K} & 
\colhead{$R_{\odot}$} & 
\colhead{$M_{\odot}$} & 
\colhead{mas} &
\colhead{$M_{\odot}/yr$} & 
\colhead{} &
\colhead{} &
\colhead{au} &
\colhead{$\mu$m} &
\colhead{mm} &
\colhead{K} &
\colhead{deg} &
\colhead{$M_{\oplus}$}
}
\startdata
 J05334160-0606074 & $3360_{-90}^{+90}$ & $0.96_{-0.10}^{+0.10}$ & $0.29_{-0.06}^{+0.06}$ & $2.43_{-0.14}^{+0.15}$ & $-8.9_{-0.6}^{+0.4}$ & $-2.4_{-0.7}^{+0.5}$ & $-1.7_{-0.8}^{+0.9}$ & $140_{-100}^{+110}$ & $2.7_{-2.4}^{+5.0}$ & $1100_{-900}^{+4000}$ & $1390_{-50}^{+50}$ & $50_{-23}^{+10}$ & $8_{-3}^{+6}$ \\
 J05340108-0602268 & $3940_{-40}^{+40}$ & $1.47_{-0.09}^{+0.09}$ & $0.67_{-0.03}^{+0.03}$ & $2.601_{-0.018}^{+0.019}$ & $-8.12_{-0.10}^{+0.11}$ & $-1.14_{-0.18}^{+0.10}$ & $-3.7_{-0.2}^{+0.3}$ & $60_{-42}^{+120}$ & $3.9_{-2.4}^{+4.0}$ & $800_{-700}^{+4300}$ & $1440_{-40}^{+40}$ & $10_{-5}^{+10}$ & $2_{-1}^{+2}$ \\
 J05340495-0623469 & $3440_{-50}^{+50}$ & $1.22_{-0.13}^{+0.11}$ & $0.33_{-0.04}^{+0.04}$ & $2.61_{-0.05}^{+0.05}$ & $-9.75_{-0.19}^{+0.40}$ & $-2.5_{-0.5}^{+1.0}$ & $-3.7_{-0.2}^{+0.5}$ & $160_{-90}^{+90}$ & $2.1_{-1.7}^{+5.0}$ & $1000_{-600}^{+3000}$ & $1420_{-50}^{+50}$ & $65_{-12}^{+3}$ & $3_{-2}^{+5}$ \\
 J05342043-0631150 & $3100_{-100}^{+100}$ & $0.89_{-0.12}^{+0.12}$ & $0.15_{-0.04}^{+0.05}$ & $2.43_{-0.15}^{+0.14}$ & $-9.7_{-0.2}^{+0.4}$ & $-3.1_{-0.3}^{+0.4}$ & $-3.7_{-0.3}^{+0.5}$ & $170_{-100}^{+90}$ & $3.2_{-2.7}^{+4.0}$ & $1000_{-700}^{+3000}$ & $1420_{-50}^{+50}$ & $57_{-20}^{+10}$ & $9_{-5}^{+11}$ \\
 J05343360-0611194 & $3020_{-30}^{+50}$ & $1.04_{-0.10}^{+0.08}$ & $0.122_{-0.011}^{+0.014}$ & $2.58_{-0.07}^{+0.07}$ & $-9.3_{-0.4}^{+0.7}$ & $-2.6_{-0.5}^{+0.6}$ & $-3.2_{-0.5}^{+0.7}$ & $160_{-100}^{+100}$ & $0.7_{-0.4}^{+4.0}$ & $1000_{-800}^{+3000}$ & $1400_{-40}^{+50}$ & $10_{-9}^{+30}$ & $5_{-2}^{+4}$ \\
\enddata
\tablecomments{This table is published in its entirety in the machine-readable format. A portion is shown here for guidance regarding its form and content.}
\end{splitdeluxetable*}

\section{Discussion} \label{sec:discussion}
Here, we discuss the importance of considering optical depth when calculating disk dust masses, particularly for massive disks, which tend to be more affected. We follow by comparing our dust mass measurements (based on SED modeling) with those from the literature \citep[based on mm fluxes;][]{2022Anderson, 2022vanTerwisga}, and show that our values become more discrepant with the values from the literature as we move toward more massive disks. Next, we compare Serpens and L1641/L1647 to 10 other star-forming regions with ages less than 10 Myr and confirm that both samples follow the observed trends for disk dust mass in other regions. We conclude by discussing how our dust masses are greater than those calculated from mm fluxes, but this does not fully solve the ``missing mass'' problem as our values are still lower when compared to exoplanet systems. 

\subsection{The importance of optical depth when deriving disk masses} \label{subsec: optical depth vs disk mass}

\begin{figure*}
    \centering
    \includegraphics[width=\textwidth]{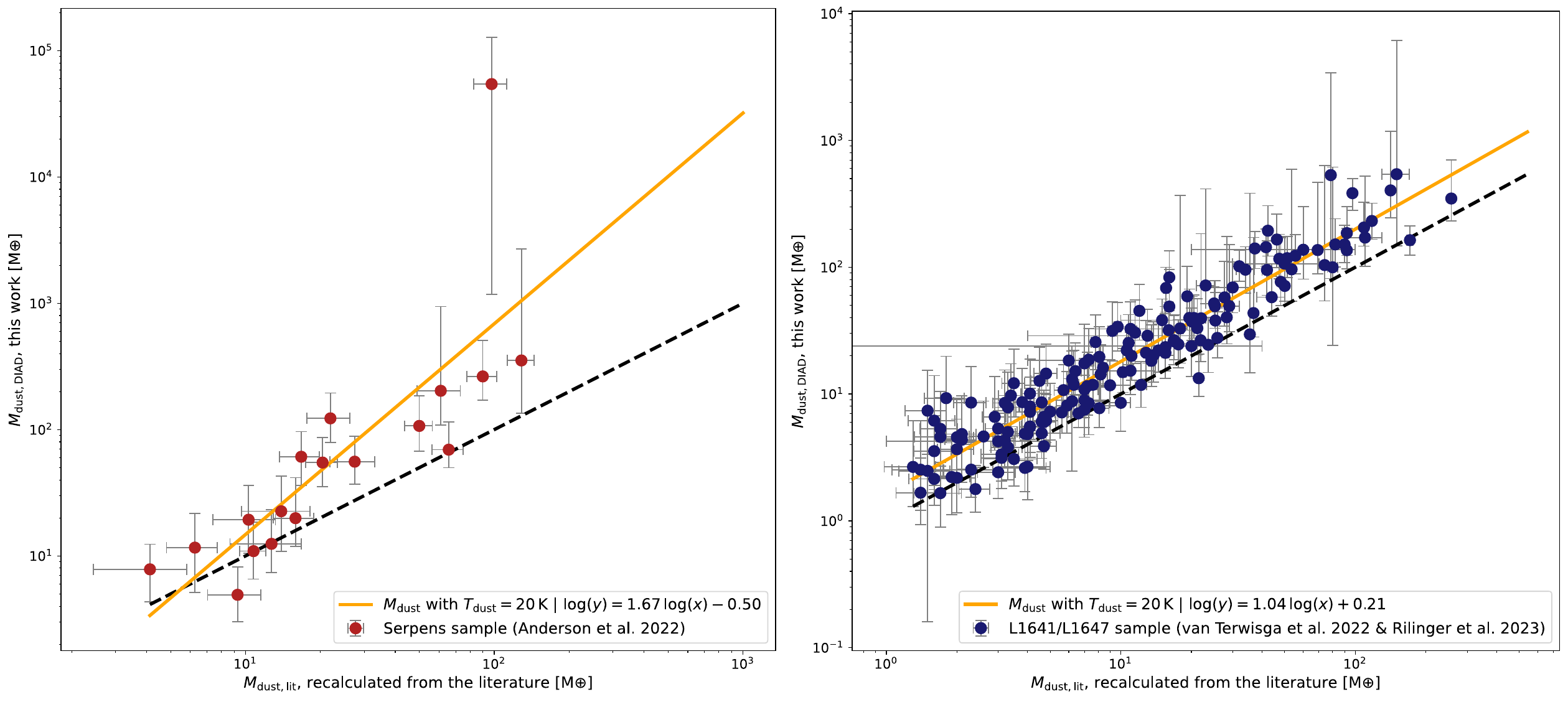}
    \caption{Comparison of this work's disk dust masses calculated from SED modeling and disk dust masses derived from millimeter fluxes from the literature; left panel corresponds to the Serpens sample  \citep{2022Anderson}, right panel corresponds to the L1641/L1647 sample \citep{2022vanTerwisga,2023Rilinger}. Both panels show a 1-to-1 relation (dotted black line) and a linear fit (solid orange line) that assumes a constant dust temperature of 20 K. The corresponding equation is noted for each linear fit.}
    \label{fig: comparison with lit}
\end{figure*}

We compare our sample's model-derived disk dust masses with those found in the literature in Figure \ref{fig: comparison with lit}. The literature values in Figure \ref{fig: comparison with lit} are from \cite{2022Anderson} \& \cite{2022vanTerwisga}, and are recalculated using \textit{Gaia} DR3 parallaxes and using the \cite{1983Hildebrand} equation: 

\begin{equation}
M_d = \frac{F_{\nu}d^2}{\kappa_{\nu}B_{\nu}(T_d)}
\label{eq: Hildebrand}
\end{equation}

where $M_d$ is the dust mass in the disk; 
$F_{\nu}$ is the millimeter flux density at the observing frequency, here 1.3 mm (241 GHz); 
$\kappa_{\nu}$ is the dust opacity coefficient at the observing frequency, and following \cite{2022Anderson}, we adopt $\kappa_{\nu} = (\nu/100$ GHz) \citep[see][]{1990Bekwith} resulting in $2.41\ cm^2g^{-1}$ while assuming $\beta = 1$; 
$B_{\nu}(T_d)$ is the Planck function evaluated at the dust temperature $T_d$, which is typically assumed to be constant with a value of $T_d = 20\ K$ for a Class II object \citep{2005AndrewsWilliams,2016Ansdell,2017Ansdell,2018RuizRodriguez,2019Cazzoletti,2019Williams,2022Anderson}; 
$d$ is the distance calculated from \textit{Gaia} DR3 parallaxes. The recalculated disk dust masses for Serpens and L1641/L1647 are listed in Tables \ref{tab: recalculated dust masses serpens} $\&$ \ref{tab: recalculated dust masses SODA}, respectively.

\begin{deluxetable}{l c}
\tablecaption{Recalculated literature dust masses for the Serpens sample. \label{tab: recalculated dust masses serpens}}
\tablehead{
\colhead{Object} &
\colhead{$M_{dust}$}
\\
\colhead{2MASS} &
\colhead{$M_{\oplus}$}
}
\startdata
J18264598-0344306 & $61 \pm 12$ \\
J18264973-0355596 & $66 \pm 9$ \\
J18274426-0329519 & $90 \pm 12$ \\
J18274445-0356016 & $10.8 \pm 1.3$ \\
J18275383-0002335 & $6.3 \pm 1.4$ \\
J18281519-0001408 & $22 \pm 4$ \\
J18290088+0029312 & $100 \pm 10$ \\
J18291969+0018030 & $27 \pm 6$ \\
J18293084+0101071 & $4.1 \pm 1.7$ \\
J18294147+0107378 & $50 \pm 6$ \\
J18295360+0117017 & $14 \pm 4$ \\
J18295820+0115216 & $130 \pm 20$ \\
J18300545+0101022 & $10.3 \pm 2.9$ \\
J18300610+0106170 & $20 \pm 3$ \\
J18302240+0120439 & $17 \pm 3$ \\
J18314416-0216182 & $9.3 \pm 2.2$ \\
J18321676-0223167 & $15.9 \pm 2.9$ \\
J18321707-0229531 & $13 \pm 4$ \\
\enddata
\end{deluxetable}

\begin{deluxetable}{l c}
\tablecaption{Recalculated literature dust masses for L1641/L1647. \label{tab: recalculated dust masses SODA}}
\tablehead{
\colhead{Object} &
\colhead{$M_{dust}$}
\\
\colhead{2MASS} &
\colhead{$M_{\oplus}$}
}
\startdata
J05334160-0606074 & $4.1 \pm 0.4$ \\
J05340108-0602268 & $1.9 \pm 0.4$ \\
J05340495-0623469 & $1.5 \pm 0.4$ \\
J05341574-0636046 & $42.0 \pm 2.0$ \\
J05342043-0631150 & $1.8 \pm 0.4$ \\
J05343360-0611194 & $2.0 \pm 0.4$ \\
J05343614-0605346 & $25.2 \pm 0.4$ \\
J05344765-0619401 & $7.6 \pm 0.3$ \\
J05344833-0630268 & $20.1 \pm 0.4$ \\
J05345404-0623507 & $82.1 \pm 0.4$ \\
\enddata
\tablenotetext{*}{This table is published in its entirety in the machine-readable format. A portion is shown here for guidance regarding its form and content.}
\end{deluxetable}

\begin{figure*}
    \centering
    \includegraphics[width=14cm]{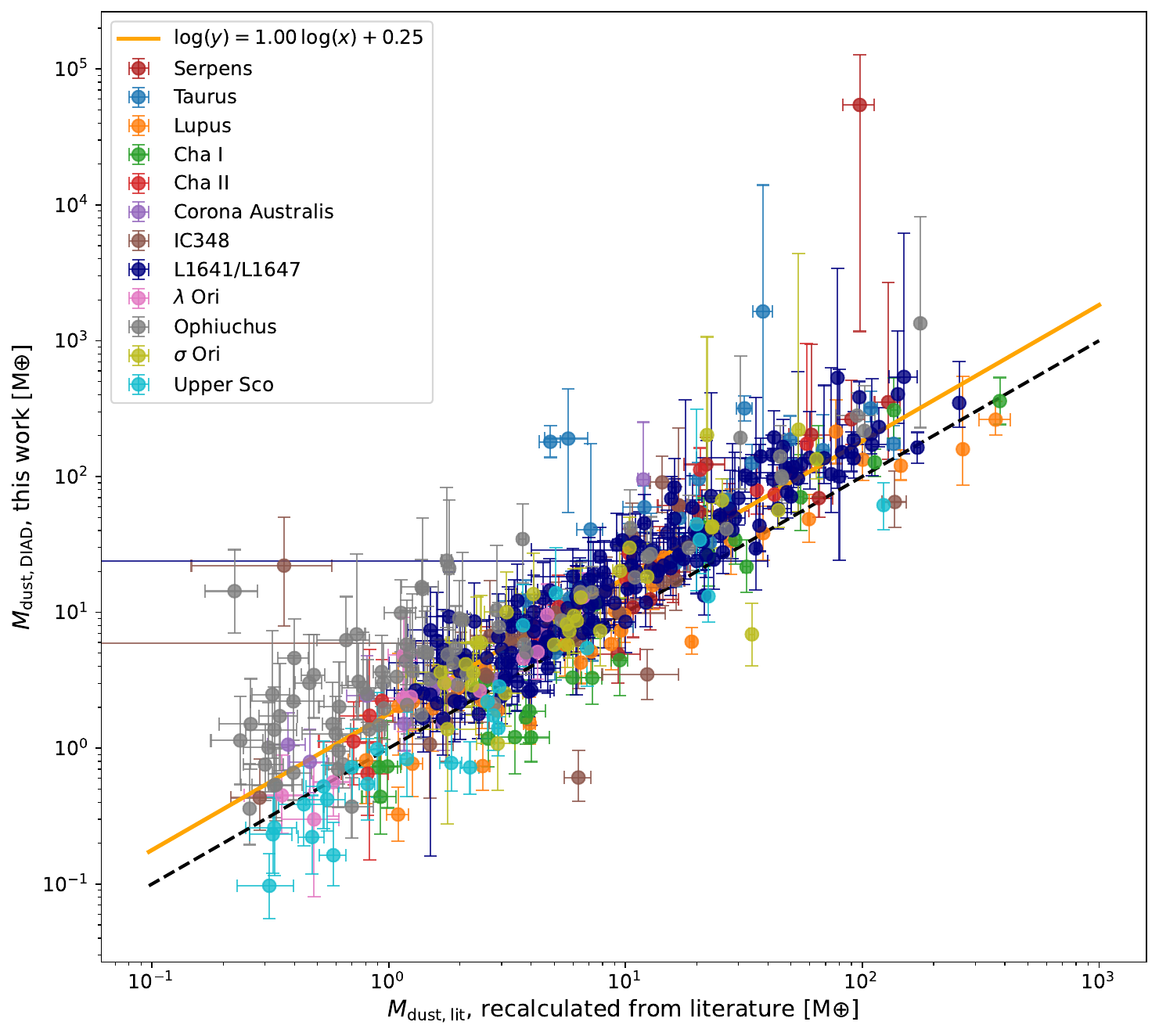}
    \caption{Comparison of this work's $M_{dust}$ calculated from SED modeling and $M_{dust}$ derived from millimeter fluxes from the literature; Serpens \citep{2022Anderson} and L1641/L1647 \citep{2022vanTerwisga} data corresponds to this work, while data for the remaining 10 star-forming regions is gathered from \cite{2023Rilinger}, where the same methodology was applied to calculate the masses. The solid orange line displays the linear fit to the recalculated dust masses, assuming a constant dust temperature of 20 K. The black dashed line denotes a 1-to-1 relation.}
    \label{fig: mdust comparison all regions}
\end{figure*}

In Figure \ref{fig: comparison with lit}, the calculated disk dust masses from the literature are consistently lower than in this work, especially when assuming a constant temperature profile. We find that, on average, our masses are a factor of about 2.0 higher for both Serpens and L1641/L1647. In most cases, the discrepancy is even higher for the higher-mass objects. 

The discrepancy between disk dust masses when using the \cite{1983Hildebrand} equation versus SED modeling has been reported in other works. \cite{2019BalleringEisner} find that the masses derived with their SED modeling are $\sim1 - 5$ higher. The same is seen in \cite{2020Ribas}, where the estimated $M_{dust}$ from SED modeling is $\sim3$ higher than that reported in \cite{2013Andrews}. Other works such as \cite{2021Macias,2022Liu} find masses $\sim3 - 5$ higher, respectively. Recent work by \citet{2023Xin,2023Rilinger} shows that the SED modeling-derived masses are, respectively, $\sim1.5 - 6\ \& \sim1.5 - 5$ times higher than those obtained from the literature. Using values from \cite{2023Rilinger} and this work for the Serpens and L1641/L1647 samples, Figure \ref{fig: mdust comparison all regions} compares all 12 regions and shows the mass ratio between disk mass derived from SED modeling and those calculated from literature fluxes.
The discrepancy factor, given as $M_{dust, DIAD} / M_{dust, lit}$, has a median of $\sim$1.74 for all regions and shows a slight increase toward the most massive disks. For younger regions (i.e., $<$2 Myr; Ophiuchus, Taurus, Cha II, L1641/L1647, Serpens, and Lupus) the discrepancy factor is greater than the populations' average (i.e., 2.0), while for older regions (i.e., $>$2 Myr; Cha I, IC 348, Corona Australis, $\sigma$ Ori, $\lambda$ Ori, and Upper Sco) the discrepancy factor is lower (i.e., 1.2). Taurus has the biggest discrepancy factor (4.09), and Upper Sco has the lowest (0.72). It is important to note that these regions have different sample sizes, so it is likely that these values are influenced by the number of objects.

The explanation that has been proposed for the discrepancy in disk masses seen above is that Equation \ref{eq: Hildebrand} assumes that the disk emission is optically thin at the reference frequency, and this may not be true in all cases. For example, there are cases where the spectral index of the inner regions drops below 2 (i.e., TW Hya \citep{2018Huang,2021Macias}, Elias 2-27 \citep{2021Paneque-Carreno}, and MP Mus \citep{2023Ribas}), indicating that the emission is optically thick. If the disk is optically thick at certain wavelengths, the observed flux will not be sensitive to the total dust mass of the disk, and so Equation \ref{eq: Hildebrand} will underestimate $M_d$. 

\begin{figure*}
    \centering
    \includegraphics[width=\textwidth]{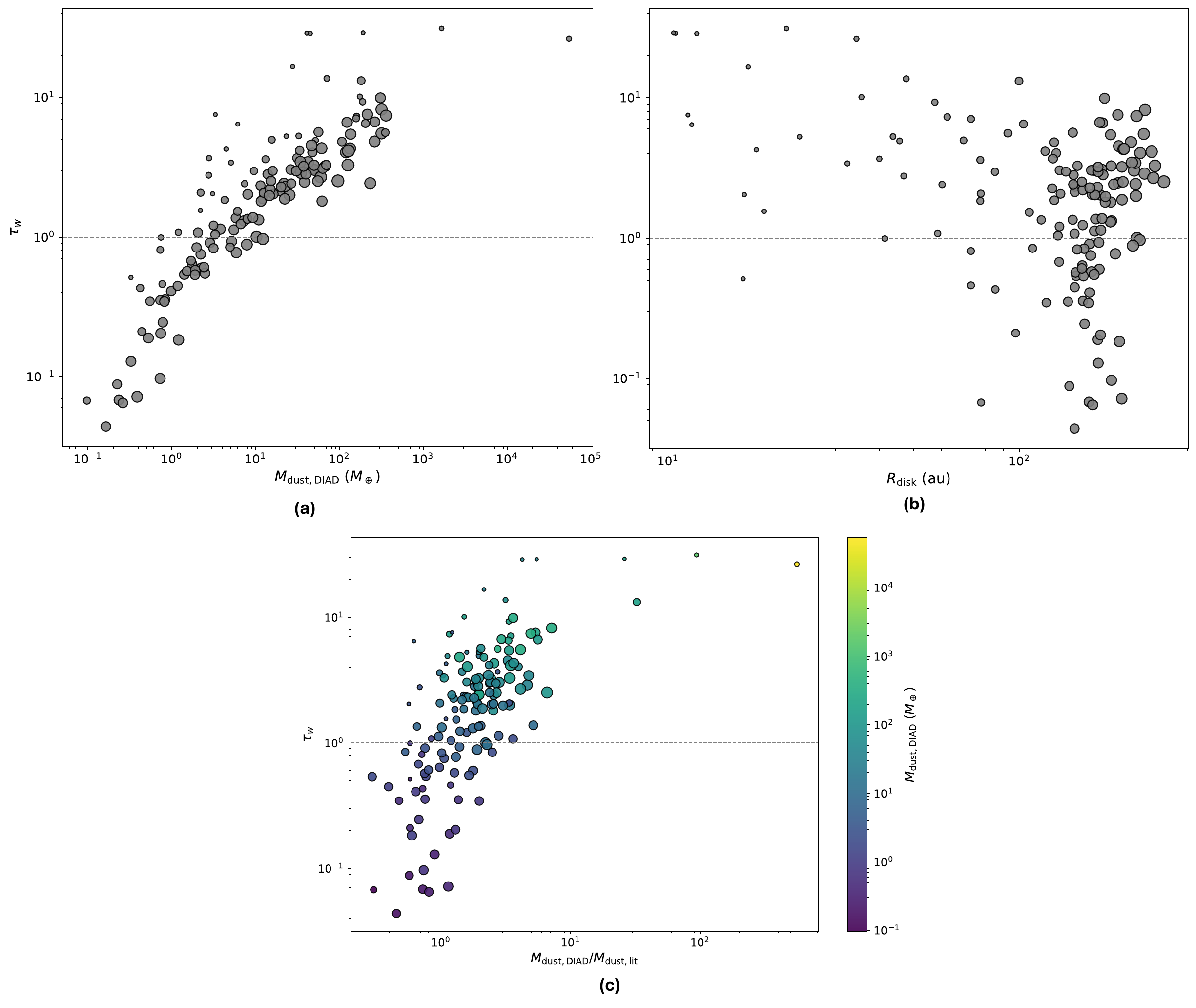}
    \caption{Comparison of flux-weighted mean optical depth for subsets of disks in different star-forming regions (i.e., L1641/L1647 [1.5 Myr], Taurus [1-2 Myr], Serpens [1-3 Myr], Lupus [1-3 Myr], Cha I [2-3 Myr], and Upper Sco [5-10 Myr]) plotted versus (a) disk dust mass, (b) disk size, and (c) disk dust mass ratio (i.e., $M_{dust, DIAD} / M_{dust, lit}$). Optical depths are obtained from flux-weighted averages of the 1.3 mm optical depth at each radius in the disk, calculated using the DIAD models. Optical depth values for the Lupus, Taurus, Cha I, and Upper Sco regions are adopted from \cite{2023Rilinger}. The denoted disk masses correspond to the median value of the distribution obtained for each object. The color bar in panel (c) indicates the dust mass of the disk. In all three panels, marker size increases proportionally with the size of $R_{disk}$) (in au), and the black broken line represents $\tau_{\omega} = 1$.}
    \label{fig: optical depth}
\end{figure*}

We demonstrate that our larger dust mass estimates – when compared with flux-based dust mass estimates – are mostly driven by high optical depths at (sub-)mm wavelengths in Figure \ref{fig: optical depth}. Following the same methodology detailed in \cite{2023Xin}, we used the median values obtained from the ANN as input parameters to generate DIAD models from which we calculate the structure of the disk and the optical depth along the line of sight as a function of the radius at 225 GHz (1.3 mm). This is followed by computing a flux-weighted mean optical depth where the flux at each radius is used as the weight for the optical depth corresponding to that radius. These weights are calculated based on the intensity, disk radius, and inclination angle, which are further computed by considering the area difference between adjacent radii and the flux at each point, accounting for the inclination angle. The single average optical depth results from the mean of the optical depths weighted by the flux-derived weights.  Panels (a) and (b) display the flux-weighted mean optical depths versus disk dust masses and disk sizes (respectively) for the samples of Taurus, Serpens, Lupus, Cha I, and Upper Sco, while for L1641/L1647, we include 10\% of the sample to avoid overpopulating the plots. The marker size indicates disk radius (in au), and the black broken line denotes $\tau_{\omega} = 1$. Panel (a) shows that as we move toward greater disk dust masses, the mean optical depth tends to increase. Panel (b) shows that a similar trend of increasing mean optical depths occurs for smaller disk sizes, since the disk total mass gets concentrated in a smaller area, hence increasing the optical depth. Panel (c) displays again the mean optical depth, plotted against a disk dust mass ratio, i.e., the disk dust mass estimated from DIAD divided by the disk dust mass estimated from the (sub-)mm flux using Equation \ref{eq: Hildebrand}. In addition, panel (c) also shows the disk dust mass derived from DIAD in the color scale, and the disk size as shown by the marker size (like in panels (a) and (b)). Panel (c) clearly demonstrates that higher disk dust mass discrepancy factors are caused by high optical depths. The three panels together show that these high optical depths are mostly driven by high disk dust masses and small disk sizes. In other words, flux-based disk dust masses are underestimated to a greater extent for disks that are more massive and/or more compact. These results highlight the importance of properly accounting for the optical depth of the disk, as high optical depths at (sub-)mm wavelengths can cause underestimation of disk mass when relying solely on flux-based estimates assuming optically thin emission.

\subsection{Comparison of Serpens $\&$ L1641/L1647 with other Star-Forming Regions} \label{subsec: comparing serpens-l1641-l1647 with other sfr}

\begin{figure*}
    \centering
    \includegraphics[width=16.5cm]{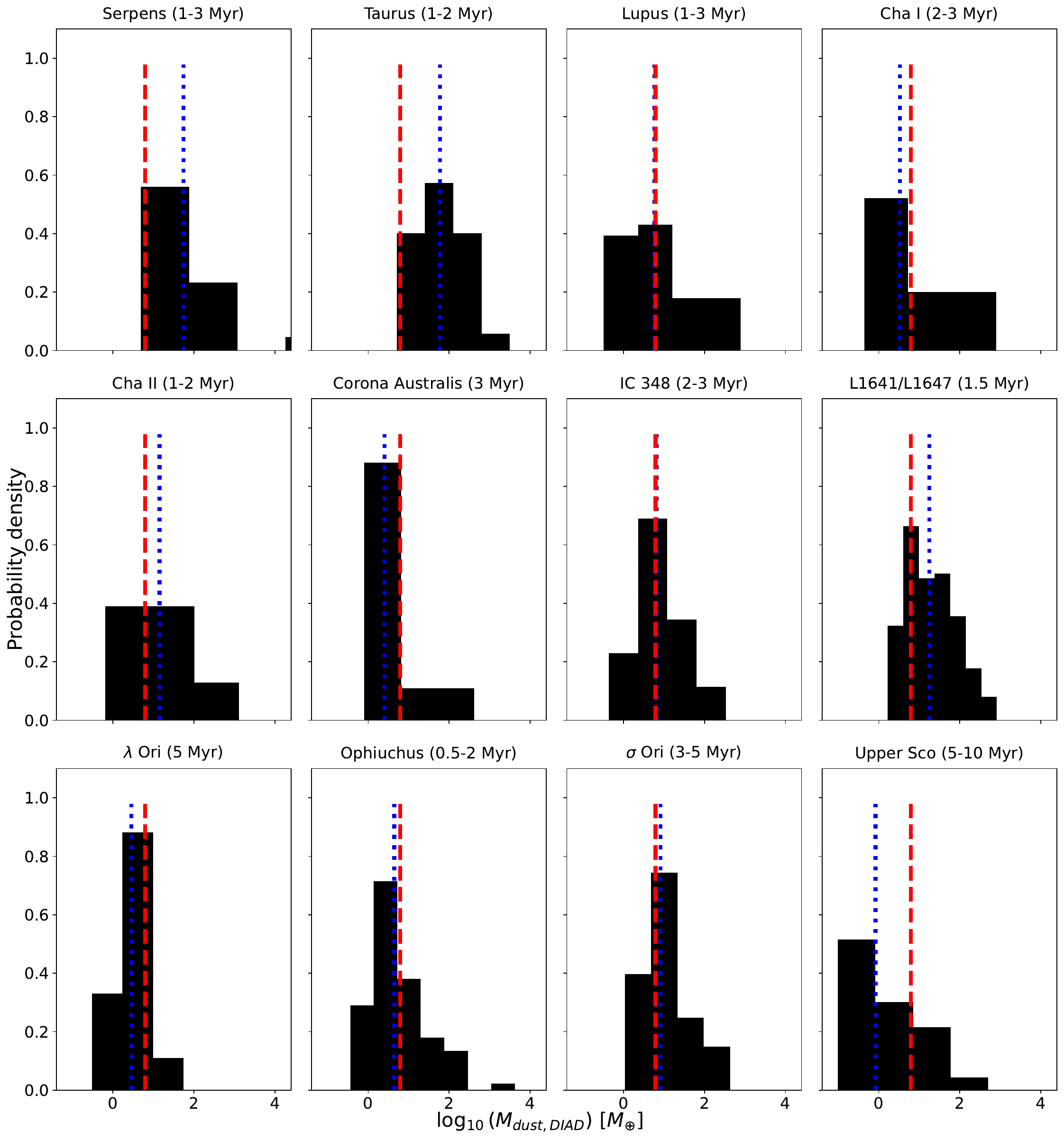}
    \caption{Probability density functions of $M_{dust}$ for 12 star forming regions. Blue dashed lines denote the median for each region, and red dotted lines denote the median for all regions ($\sim$9.28 $M_{\oplus}$). Serpens and L1641/L1647 data correspond to this work, while data for the remaining star-forming regions is gathered from \cite{2023Rilinger}. }
    \label{fig: comparison with SFR}
\end{figure*}

\begin{figure*}
    \centering
    \includegraphics[width=13cm]{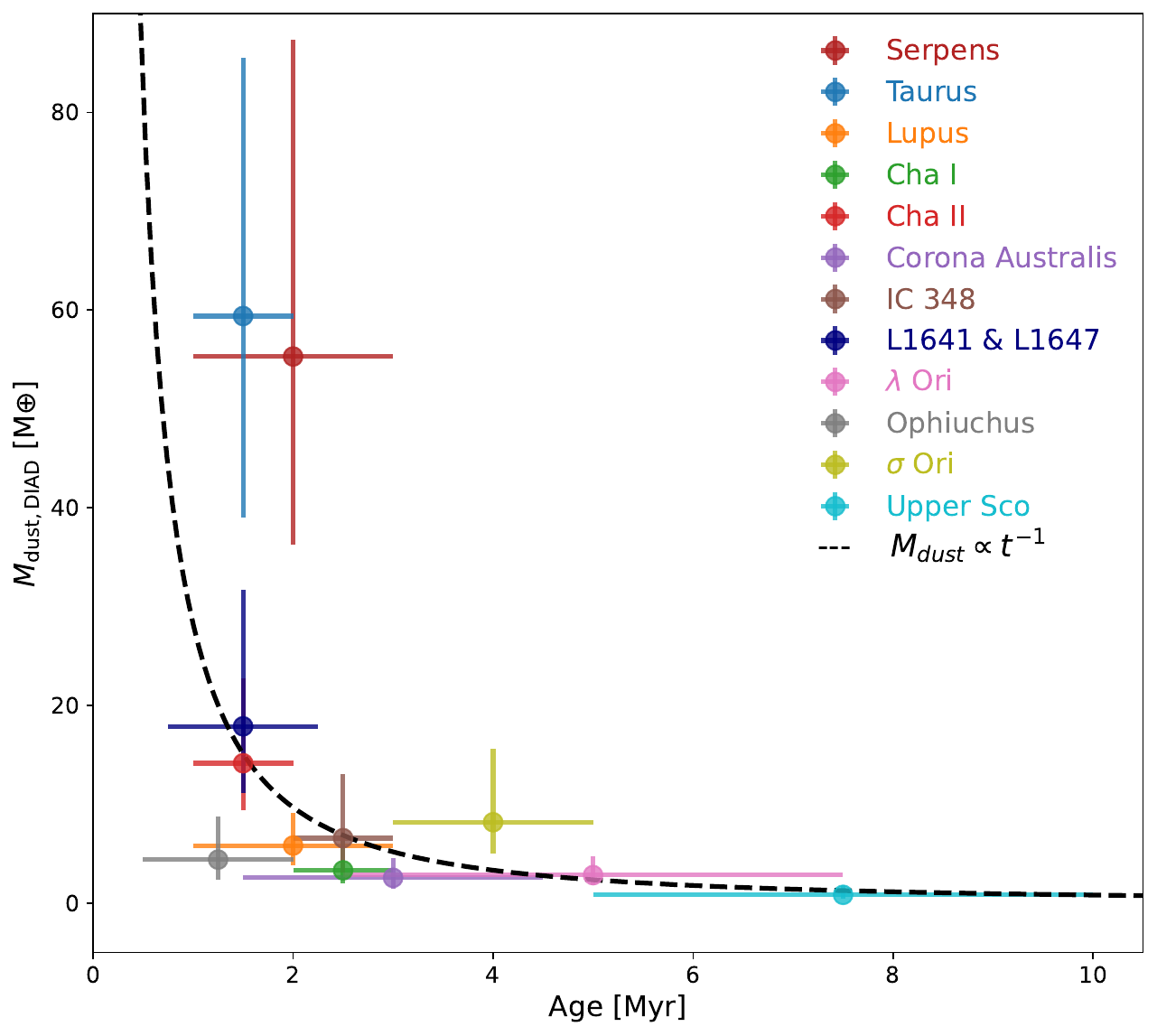}
    \caption{Median dust masses for 12 star-forming regions compared to the respective age of each region. Following \cite{2023Rilinger}, the black dashed line denotes the dust depletion function $M_{dust} \propto t^{-1}$ \citep{2022Testi}, and scales it by the weighted mean mass at 1.5 Myr. Median values for M$_{dust}$ are calculated for each region. Serpens data correspond to this work, while data for the remaining star-forming regions is gathered from \cite{2023Rilinger}.}
    \label{fig: dust depletion}
\end{figure*}

Here, we compare the results obtained for Serpens and L1641/L1647 with the 10 other star-forming regions studied in \cite{2023Rilinger}. These regions span an age range from $\sim$0.5-10 Myr.\footnote{We remove the following eight objects from the Lupus sample adopted in \cite{2023Rilinger} due to their uncertain membership: 2MASS J15450634-3417378 and 2MASS J16075475-3915446 are not detected in \textit{Gaia} DR3; 2MASS J16070384-3911113 is detected in \textit{Gaia} DR3 but has no parallax measurement; 2MASS J15560210-3655282, 2MASS J16030548-4018254, and 2MASS J16124373-3815031 have kinematics consistent with Lupus, but also Sco-Cen \cite{2020Luhman}, making their membership unlikely; 2MASS J15592523-4235066 has kinematics inconsistent with Lupus and instead indicative of  Upper Centaurus–Lupus and Lower
Centaurus–Crux; and Sz 90, which has \textit{Gaia} DR2 astrometry that supports membership in Lupus, but \textit{Gaia} DR3 data does not support it.} 
Figure \ref{fig: comparison with SFR} shows how the Serpens and L1641/L1647 dust mass distributions compare to the star-forming regions studied in \cite{2023Rilinger}. The peak of the mass distribution tends to shift to lower masses for older regions. Regions such as Serpens, Taurus, and L1641/L1647 (i.e., $<$ 3 Myr) have median dust masses above the overall median ($\sim$9.28 $M_{\oplus}$), while older regions such as Upper Sco and $\lambda$ Ori  (i.e., $>$ 3 Myr) have dust masses that are below the overall median. Figure \ref{fig: dust depletion} is also consistent with the finding that disk dust mass generally decreases with age \citep[e.g.,][]{2016Ansdell,2016Barenfeld,2016Pascucci,2019Cieza,2019vanTerwisga,2021Villenave,2022vanTerwisga,2023Rilinger}. We note that the objects included in each region are biased towards higher masses as we only include sources with millimeter-wavelength detections. Additionally, each survey adopts a different detection limit \citep{2022vanTerwisga,2022Anderson,2023Rilinger}.

\subsection{The ``missing mass'' problem}
One possible solution to the ``missing mass'' problem is that disk dust masses are underestimated.  In this work, as in other works \citep{2019BalleringEisner,2020Ribas,2023Xin,2023Rilinger}, we find that disk dust masses can be underestimated when using millimeter fluxes alone. To further explore whether this can solve the ``missing mass'' problem, we compare observed exoplanet systems and the available dust mass in protoplanetary disks. 

\begin{figure*}
    \centering
    \includegraphics[width=14cm]{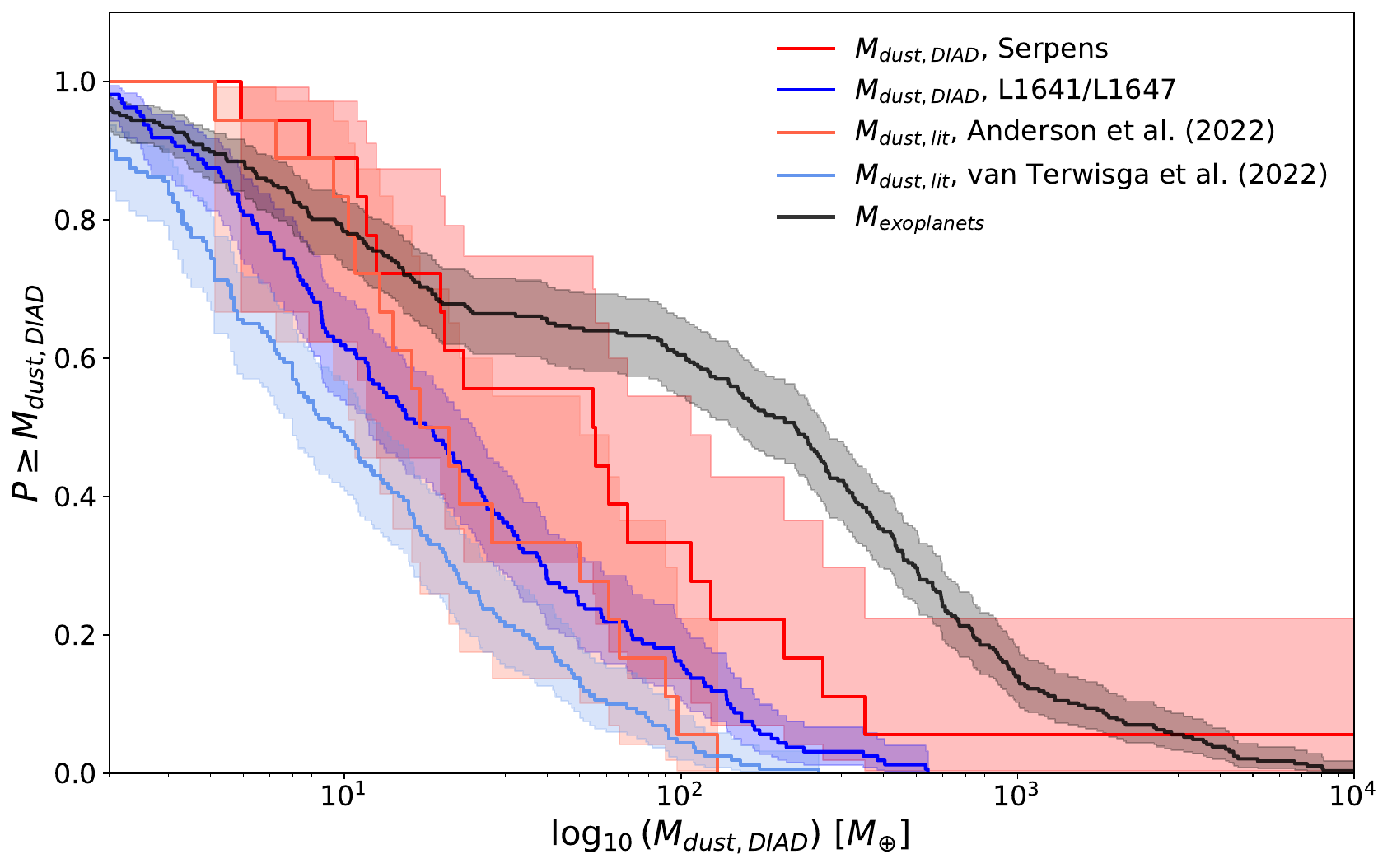}
    \caption{Comparison of cumulative distribution functions which correspond to dust masses of Serpens (red line) and L1641/L1647 (blue line) gathered from this work, and Serpens' (light red line) and L1641/L1647's (light blue line) dust masses recalculated using literature millimeter fluxes \citep{2022Anderson,2022vanTerwisga} in Equation \ref{eq: Hildebrand}, and mass content available in confirmed exoplanets which are above the detectability threshold (i.e., $>$ 0.3 $M_{Jup}$), using data gathered from the \href{https://exoplanetarchive.ipac.caltech.edu/docs/counts_detail.html}{NASA Exoplanet Archive} (black line). The cumulative distribution functions were developed using the Kaplan-Meier fitting method.}
    \label{fig: CDF}
\end{figure*}

Figure \ref{fig: CDF} compares the cumulative distribution function (CDF) for the dust mass content in Serpens and L1641/L1647 from this work (i.e., SED model derived; red line and blue line, respectively), the dust mass content using \cite{2022Anderson} \& \cite{2022vanTerwisga} millimeter fluxes (light red line and light blue line, respectively), and the mass content observed in currently known exoplanets (gray line). The masses of presently known and confirmed exoplanets were gathered from the  \href{https://exoplanetarchive.ipac.caltech.edu/docs/counts_detail.html}{NASA Exoplanet Archive}, amounting to a sample of 3401 exoplanets with known masses and respective errors. The black cumulative distribution function denotes the masses of exoplanet systems, which are normalized to the observed fraction of gaseous planets \citep{2008Cumming}. Based on observations, gaseous planets are rarer than rocky low-mass planets. As shown in \cite{2011Mayor}, about 50$\%$ of solar-type stars host planets with a mass upper limit $<$ 30 M$_{\oplus}$. \cite{2008Cumming} estimates that about 17$\%$-19$\%$ of stars host gas giants within 20 au. Following the methodology detailed in \cite{2020Tychoniec}, we restrict our sample to only systems containing at least one planet with a total mass $>$ 0.3 $M_{Jup}$, as detectability increases for objects with periods as large as 2000 days \citep{2008Cumming}. We normalize these systems by setting the 18$\%$ value of the CDF at the estimated solid fraction of the gaseous planet of 0.3 $M_{Jup}$ (i.e., 27.8 $M_{\oplus}$).
Thus, we limit our sample to only those objects that are detectable and have a known solid material content \citep[were the solid mass for gaseous planets is estimated using the formula from][]{2016Thorngren}. Figure \ref{fig: CDF} shows how, despite accounting for disk mass underestimation, the ``missing mass" issue persists, implying that the problem is more complicated and may involve multiple factors.

As mentioned in Section \ref{sec:intro}, the ``lack" of mass can also be explained by earlier planet formation (i.e., happening as early as the Class 0 and Class I stages), as discussed in \cite{2020Tychoniec} where they compare Class 0, Class I, and Class II objects from the Perseus and Orion star-forming regions. In both instances, the CDF for Class II objects does not follow the CDF for exoplanets at greater masses. If planet formation occurs earlier (i.e., during the Class O and/or Class I stages), then the dust masses we are measuring for Class II objects may only represent that of the remnant material of planet formation.

\begin{figure*}
    \centering
    \includegraphics[width=15cm]{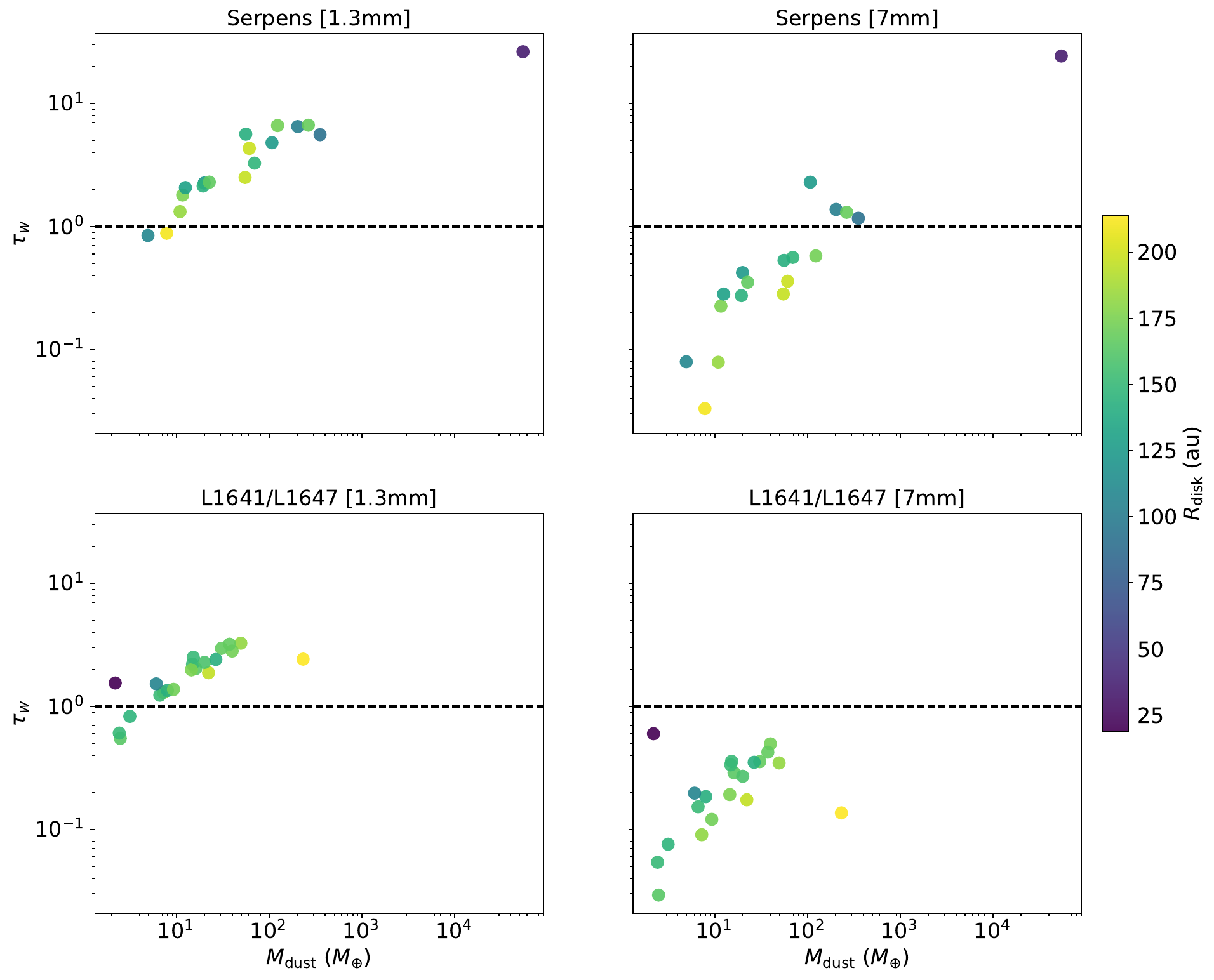}
    \caption{Comparison of flux-weighted mean optical depth for a subset of disks in Serpens (top) and L1641/L1647 (bottom) plotted versus disk dust mass. Optical depths are gathered from flux-weighted averages of the 1.3 mm (left) and 7 mm (right) optical depth at each radius in the disk, calculated using the DIAD models. Denoted disk masses correspond to the median value of the distribution obtained for each object. We note that for Serpens we show the whole sample of 18 objects, while for L1641/L1647 we show the 10$\%$ of the sample. The color bar denotes the disk's outer radius (in au). The black broken line represents $\tau_{\omega} = 1$.}
    \label{fig: optical_depth_7mm}
\end{figure*}

It is also possible that we are still underestimating the dust masses.  First, we are not sensitive to grains $> 1$~mm. In addition, there may be significant dust in optically thick disk substructures \citep{2024Savvidou}. To more accurately measure disk dust masses, observations at longer wavelengths are necessary. \cite{2023Xin} showed that disks become optically thinner at 3 mm relative to 1.3 mm. In Figure~\ref{fig: optical_depth_7mm}, we calculate the optical depth at a longer reference wavelength of 7mm corresponding to ALMA band 1 for Serpens and L1641/L1647. We find that the vast majority of objects become optically thin at 7 mm. Thus, observations at longer mm wavelengths (i.e., ALMA band 3 and band 1) are essential to better characterize the dust content of disks.

\section{Summary}
\label{sec:summary}
We calculated disk dust masses for a sample of 178 TTS, 18 located in the Serpens star-forming region and 160 in the L1641 \& L1647 regions, using multi-wavelength photometry and \textit{Spitzer} IRS spectra (when available) and an ANN trained with DIAD models (following \cite{2020Ribas}). The sample was obtained from ALMA 1.3 mm continuum band surveys of Serpens \citep{2022Anderson} and the L1641 \& L1647 regions of the Orion A cloud \citep{2022vanTerwisga}.
Our work measured disk dust masses using SED modeling and the key findings are as follows: 

\begin{itemize}
    \item Disk dust masses calculated from SED modeling are, on average, a factor of 2.0 times higher for both Serpens and L1641/L1647 than those computed using mm-fluxes at 1.3mm.   
    This is consistent with previous work done in other regions by \cite{2020Ribas,2023Xin,2023Rilinger}, where a similar methodology was followed.  We conclude that this is because most disk dust masses calculated using mm-fluxes assume that the disk is optically thin, whereas we find that many disks are still optically thick at 1.3mm.
    \item  When compared to 10 other star-forming regions that range from $\sim$0.5 to $\sim$10 Myr \citep{2023Rilinger}, Serpens and L1641/L1647 follow the tendency of younger regions (e.g., Taurus, Cha II, $\sigma$ Ori) to display a greater median disk dust mass than the overall average. Serpens' median disk dust mass is $\sim$55.3 $M_{\oplus}$, and L1641/L1647's median disk dust mass is $\sim$17.9 $M_{\oplus}$; whereas the median for all 12 star-forming regions is $\sim$9.28 $M_{\oplus}$. 
    \item Dust masses for Serpens and L1641/L1647 follow the observed relation between dust mass and age (i.e., $M_{dust} \propto t^{-1}$).
    \item We calculated the optical depths of a subset of disks in Serpens and L1641/L1647 at 1.3mm and 7mm and found that most disks are optically thick at 1.3mm while the vast majority are optically thin at 7 mm.  
\end{itemize}

Future ALMA surveys are needed at longer wavelengths where the disk is more likely to be optically thin (e.g., band 1) to more accurately measure disk dust masses.  
This will allow for a more detailed view of disk evolution and planet formation as well as comparison with exoplanet populations.

\section*{Acknowledgments}
This work was funded by NASA ADAP 80NSSC20K0451. Álvaro Ribas has been supported by the UK Science and Technology Facilities Council (STFC) via the consolidated grant ST/W000997/1 and by the European Union’s Horizon 2020 research and innovation programme under the Marie Sklodowska-Curie grant agreement No. 823823 (RISE DUSTBUSTERS project). We thank Alexa R. Anderson, Jesús Hernández, Kevin Luhman, Anneliese M. Rilinger, and Zihua Xin for insightful discussions. This paper utilizes the D’Alessio Irradiated Accretion Disk (DIAD) code. We acknowledge the important work carried out by Paola D’Alessio, who passed away in 2013. Without her, works like these could not have been achieved.

This paper makes use of the following ALMA data: ADS/JAO.ALMA\#2019.1.00218.S and ADS/JAO.ALMA\#2019.1.01813.S. ALMA is a partnership of ESO (representing its member states), NSF (USA) and NINS (Japan), together with NRC (Canada), NSTC and ASIAA (Taiwan), and KASI (Republic of Korea), in cooperation with the Republic of Chile. The Joint ALMA Observatory is operated by ESO, AUI/NRAO and NAOJ. The National Radio Astronomy Observatory is a facility of the National Science Foundation operated under cooperative agreement by Associated Universities, Inc.

This research has made use of the NASA Exoplanet Archive, which is operated by the California Institute of Technology, under contract with the National Aeronautics and Space Administration under the Exoplanet Exploration Program. 
This work has made use of the interactive graphical viewer and editor for tabular data TOPCAT \citep{2005TOPCAT}.
This research has made use of the VizieR catalogue access tool, CDS, Strasbourg, France (DOI : 10.26093/cds/vizier). The original description of the VizieR service was published in \cite{2000Vizier}.
This research has made use of the SIMBAD database \citep{2000SIMBAD}, operated at CDS, Strasbourg, France.
This work has made use of data from the European Space Agency (ESA) mission \href{https://www.cosmos.esa.int/gaia}{\textit{Gaia}}, processed by the \href{https://www.cosmos.esa.int/web/gaia/dpac/consortium}{\textit{Gaia} Data
Processing and Analysis Consortium} (DPAC). 

\vspace{5mm}
\software{\texttt{AstroPy} \citep{2013Astropy,2018AJ....156..123A,2022ApJ...935..167A}, \texttt{NumPy} \citep{2020Numpy}, \texttt{SciPy} \citep{2020Scipy}, \texttt{lifelines} \citep{2022lifelines}, \texttt{matplotlib} \citep{2007matplotlib}, \texttt{pandas} \citep{2022pandas}}

\bibliography{new.ms}{}
\bibliographystyle{aasjournal}

\appendix
\section{Sample selection criteria}
\label{sec:appendixA}
The original samples of 320 Serpens sources \citep{2022Anderson} and 873 L1641/L1647 sources \citep{2022vanTerwisga} are reduced to samples of 23 and 188 objects, respectively, after considering the following selection criteria: 

\begin{itemize}
  \item The sources must belong to the Serpens or L1641/L1647 star-forming regions. 
  
  \cite{2022Anderson} selected their sources by inspecting the \textit{Spitzer} c2d and Gould Belt surveys, which consisted of 1546 YSOs identified towards the Serpens region. They selected a sample of 320 objects for their ALMA survey by choosing only those sources with infrared slopes less than 1.6 \citep[representative of Class I, Flat, and Class II YSOs;][]{2011WilliamsCieza}. In their observation results table \citep[Table 1;][]{2022Anderson}, the \textbf{G} flag is employed to denote sources that are situated beyond the probable distance range of Serpens members. This reduces the reported observed sample from 320 to 302 likely members of the Serpens region. For further information regarding the process of membership selection, we refer the reader to \cite{2015Dunham} and \cite{2022Anderson}.
  
  The SODA sample from \cite{2022vanTerwisga} consists of 873 objects that belong to the L1641 and L1647 regions, located in the Orion A cloud at $<$ -6$\degree$ deg in declination. These were selected from the {\it Spitzer} survey of YSOs in the Orion A and B molecular clouds performed by \cite{2012Megeath}. The 41 objects in L1641, already modeled in \cite{2023Rilinger}, are included in our sample (see Section \ref{subsec: comparing serpens-l1641-l1647 with other sfr}).  
  For consistency, we use the same source identifier as in \cite{2012Megeath} and \cite{2022vanTerwisga}, abbreviating with [MGM2012] and the number of the object in the catalog.

  \item The sources must be identified as Class II. 
  
  For the Serpens sample, we used the YSO class flag included in the \cite{2022Anderson} results table \citep[based on the infrared slope value characteristic of a Class II object;][]{2011WilliamsCieza}, reducing the sample to 235 objects. The \cite{2022vanTerwisga} sample is comprised entirely of Class II YSOs from \cite{2012Megeath}, so no sources were discarded from the L1641/L1647 sample based on this criterion. 

  \item The sources must have been detected by ALMA.
  
  Non-detections were included in the \cite{2022Anderson} table for completeness. Discarding these sources reduces the sample to 182 objects, while for the L1641/L1647 sample, we only selected resolved sources, reducing the number of objects to 502. 
  
  \item The sources must have at least two 2MASS measurements, two mid-IR (\textit{WISE} or cd2) measurements, and one (sub)mm (ALMA) measurement.
  
  A photometric point at $\geq$ 500$\mu$m ensures coverage up to the (sub)mm regime, ideal for constraining the disk mass. As our models have many parameters, we require a minimum number of photometric points to obtain an appropriate fit and prevent degeneracies at the moment of SED modeling. 2MASS provides information on stellar properties, while the mid-IR traces the disk's emission. Applying this criterion does not alter the size of the Serpens sample, as all sources have measurements; meanwhile, the L1641/L1647 sample is affected and reduced to 442 objects.
  
  \item The source is not known to be a transitional disk (TD),  have a flat spectrum, or be in a binary/multiple system.

  The ANN omits TDs in its training sample \citep{2020Ribas}. Binary and multiple systems are avoided since they cause confusion when compiling their SED data, and in cases of low spatial resolution, this leads to inaccurate photometry. We discarded sources with companions $<$1000 au, based on the literature, as binarity can impact protoplanetary disk properties, structure, and evolution \citep[e.g.,][]{1994Artymowicz,2009Cieza,2012Kraus,2012Harris,2016Kounkel,2016RuizRodriguez,2017Cox,2019Akeson,2019Barenfeld,2019Cazzoletti,2021Villenave,2022Rota}. We refer the reader to \cite{2022Offner}, for a recent review on the effects of binarity on disk systems.

  For the Serpens sample, we identified objects as TDs following \cite{2016vanderMarel}, and used the provided flag \textbf{T} in \cite{2022Anderson}, or used visual inspection of the SED to identify a ``dip'' in the mid-IR range (e.g., 2MASS J18300893+0047219 shows a SED characteristic of a TD). This reduces the Serpens sample to 163 objects. 
  
  For the L1641/L1647 sample, we consulted the \cite{2013Kim} infrared spectroscopic survey of Orion A on transitional disks, finding 19 objects in our sample identified as a type of TD (ranging from classical TD, weak-excess TD, and pre-TD). From the \cite{2013FangA} L1641 study, we identified 23 objects as TD or Flat spectrum. Visually we identified and discarded [MGM2012] 128, 828, and 1066 as they show a ``dip'' around the mid-IR range, characteristic of the SED structure of a TD. From the \cite{2009Fang} catalog, six objects are identified as TD: [MGM2012] 339, 365, 377, 395, 481, 971. We exclude five objects associated with extended millimeter-continuum emission from the envelope ([MGM2012] 56, 227, 421, 714, and 975; \cite{2022vanTerwisga}). We also discard compact (visual) systems with companions within $<$10" (4000 au) (e.g., [MGM2012] 129, 754, 1000, 1090, and 1094). This reduces the L1641/L1647 sample to 381 objects. 
  
  \item The sources must not have a (sub)mm measurement within 2''.5 of another (sub)mm detection of at least 3$\sigma$. 
  
  For the Serpens sample, we used the provided flag \textbf{B} in \cite{2022Anderson}, which indicates that an object is detected in a binary system, with a 3$\sigma$ mm detection within 2''.5 ($\sim$1000 au) of the central source. This reduces the sample to 162 after discarding 2MASS J18311972-0201115. For the L1641/L1647 sample, we discarded one class II source ([MGM2012] 512) associated with extended millimeter-continuum emission \citep{2021Grant,2024Cacciapuoti}, reducing the sample to 380 objects. 

  \item The sources must have an ALMA detection of at least 3$\sigma$
  
  This reduces the Serpens sample to 70 objects while it keeps the L1641/L1647 sample unchanged.
  
  \item The sources must display good \textit{Gaia} astrometric quality.
  
  To satisfy this, we select sources with \textit{Gaia} DR3 astrometry. 
  This reduces the Serpens sample to 42 sources. Although 2MASS 18314182-0207590 is included as a \textit{Gaia} DR3 object, it is discarded as it does not have parallax values. We follow \citet{2021Lindegren}, which recommends the use of the renormalized unit weight error (RUWE, the uncertainty in measurements) to evaluate the quality of the astrometric solution. We apply two criteria for good astrometric fits: a parallax error below 20$\%$, and a RUWE $<$1.4. 
  This reduces the Serpens sample to 29 sources, while the L1641/L1647 sample is reduced to 205 objects. 

  \item The sources must be T Tauri stars.
  
  The ANN training is optimized with DIAD models that cover an estimated stellar mass in the range from 0.1 - 2.0 $M_{\odot}$ \citep{2020Ribas}.  This generally corresponds to the range of masses associated with T Tauri stars, and so we discard objects classified as Herbig Ae/Be stars to ensure that no stars outside the stellar mass range covered by the ANN are included. From the Serpens sample, 2MASS J18295533+0049391 was discarded as it is identified as a Herbig Ae/Be star in \cite{2009Oliveira}. This results in the reduction of the Serpens sample to 28 objects. 
  In the L1641/L1647 sample, we identified four sources as Herbig Ae/Be following \cite{2021Guzman-Diaz}: [MGM2012] 612, 760, 838, 1031. This reduces the L1641/L1647 sample to 201 objects. 

  \item The sources must have stellar parameters in the literature.
  
  Stellar parameters are listed in Tables \ref{table:parameters_serpens} \& \ref{table:parameters_SODA}. For Serpens, we use \cite{2007Harvey, 2009Oliveira,2013Oliveira,2015Erickson} for the star's mass, radius, effective temperature, spectral type, and visual extinction. 
  For the L1641/L1647 sources, we use \cite{2009Fang,2013FangA,2018Fang,2012Hsu,2013Hsu,2016DaRio,2016Kounkel,2016Kim,2018Yao,2023Zuniga,2023Serna,2023Abdurro}.
  Tables \ref{table:parameters_serpens} \& \ref{table:parameters_SODA} also include parallaxes and their respective error values obtained from \textit{Gaia} DR3. These are all parameters later used as priors for the ANN. 
  
  After inspecting all objects, we found that 5 objects from Serpens did not have parameters available in any catalog: [2MASS] 18274937-0353404, 18314002-0205339, 18314616-0153493, 18314791-0211559, 18325715-0245162. For L1641/L1647, 13 objects do not have available parameters: [MGM 2012] 102, 108, 121, 132, 146, 152, 168, 171, 229, 249, 260, 683, 974. These sources were discarded, reducing the Serpens sample to 23 and the L1641/L1647 sample to 188. 
\end{itemize}

\section{Priors vs Posteriors}
\label{sec:appendixB}
\begin{figure*}
    \centering
    \includegraphics[width=\textwidth]{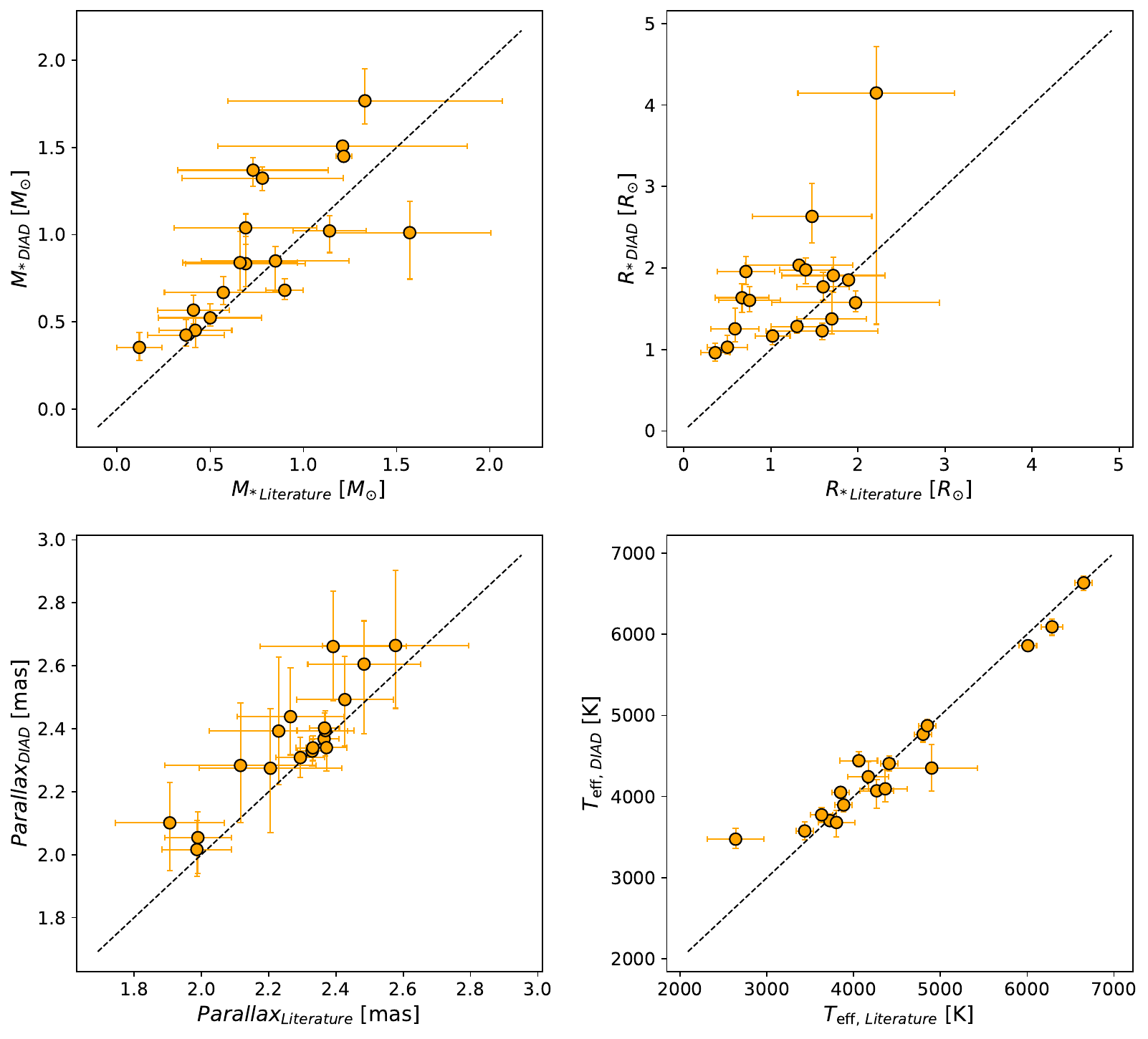}
    \caption{Comparison between the existing literature prior values and the posterior value estimations derived from the ANN analysis of the Serpens disks. The dotted black line denotes a perfect correlation (one-to-one fit) between the values.}
    \label{fig: serpens_posteriors_priors}
\end{figure*}

\begin{figure*}
    \centering
    \includegraphics[width=\textwidth]{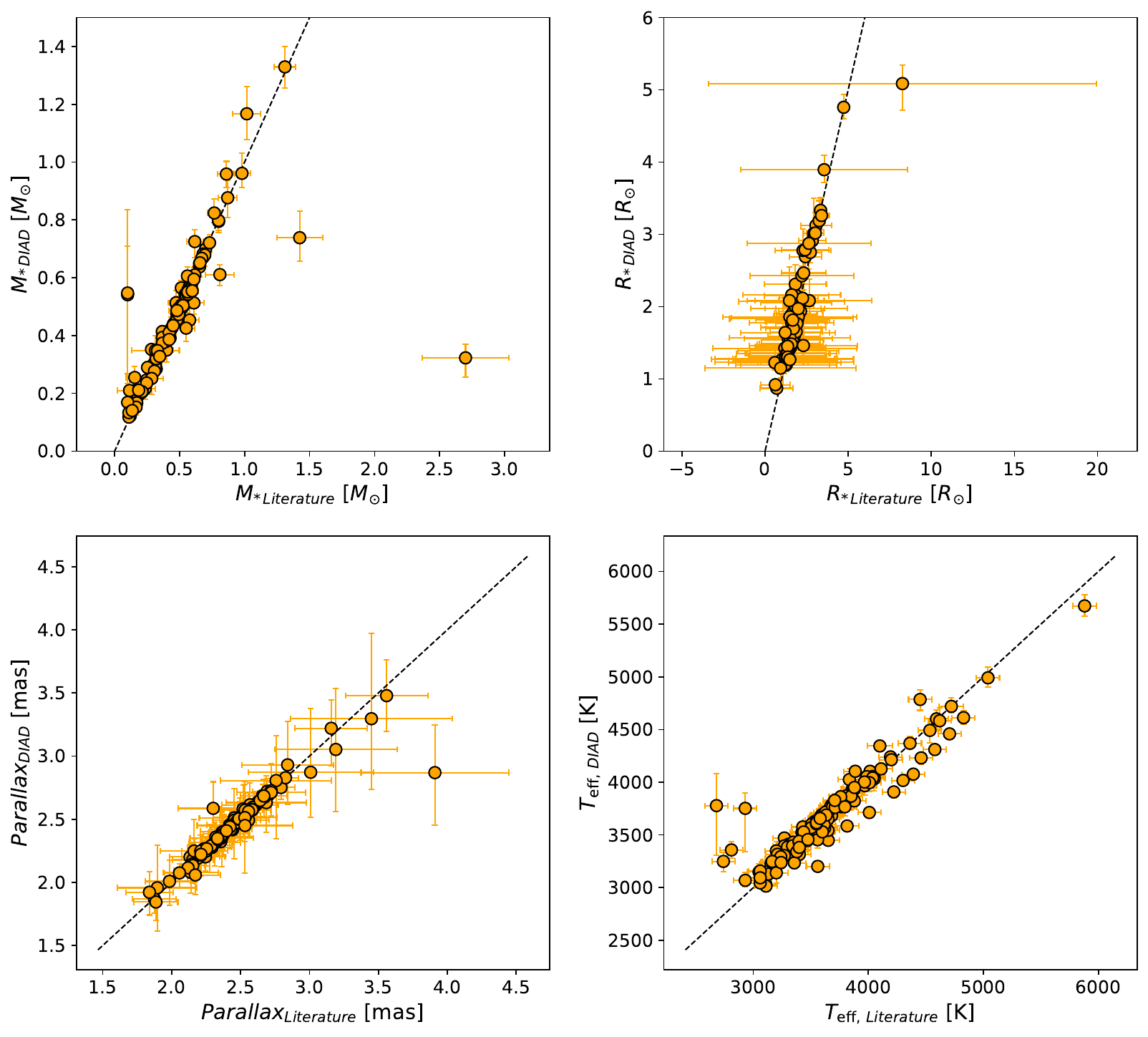}
    \caption{Comparison between the existing literature prior values and the posterior value estimations derived from the ANN analysis of the L1641 \& L1647 disks. The dotted black line denotes a perfect correlation (one-to-one fit) between the values.}
    \label{fig: L1641_L1647_posteriors_priors}
\end{figure*}

An evaluation of the consistency between the stellar parameters $M_*$, $R_*$, parallax, and $T_{\mathrm{eff}}$ (as listed in Tables \ref{table:parameters_serpens} and \ref{table:parameters_SODA} and used as priors in our modeling) and their corresponding posterior distributions (in Tables \ref{table: ANN_posteriors_Serpens_stellar} and \ref{tab: ANN_posteriors_SODA_stellar}) is performed to check for sources where our modeling suggests significant discrepancies with literature measurements. Figures \ref{fig: serpens_posteriors_priors} and \ref{fig: L1641_L1647_posteriors_priors} illustrate the comparison between existing literature measurements and the posterior probability distributions obtained via the ANN. As expected, the posterior distribution for the disks' stellar parameters are in general consistent with literature measurements. This suggests that the literature values used as priors are not problematic, and our methodology allows us to incorporate the uncertainty on these stellar parameters in the posterior distributions of the disk parameters \citep{2020Ribas}.

\end{document}